\documentclass[a4paper,fleqn,usenatbib]{mnras}

\usepackage{newtxtext,newtxmath}
\usepackage[T1]{fontenc}
\usepackage{ae,aecompl}
\usepackage{graphicx}	% Including figure files
\usepackage{amsmath}	% Advanced maths commands
\usepackage{amssymb}	% Extra maths symbols
\usepackage{hyperref}
\usepackage{longtable}
\usepackage{multicol}
\usepackage[flushleft]{threeparttable}
\usepackage{pdflscape}
\title[]{The temporal evolution of neutron-capture elements in the Galactic discs}

\author[Spina et al.]
{Lorenzo Spina,$^{1,2}$\thanks{E-mail: lspina@monash.edu}
Jorge Mel\'endez,$^{1}$
Amanda I. Karakas,$^{2}$
Leonardo dos Santos,$^{1,3}$\newauthor
Megan Bedell,$^{4,5}$
Martin Asplund,$^{6}$
Ivan Ram\'irez,$^{7}$
David Yong,$^{6}$
Alan Alves-Brito,$^{8}$\newauthor
Jacob L. Bean,$^{4}$
and Stefan Dreizler$^{9}$
\\
$^{1}$Universidade de S\~ao Paulo, IAG, Departamento de Astronomia, Rua do Mat\~ao 1226, S\~ao Paulo, 05509-900 SP, Brasil\\
$^{2}$Monash Centre for Astrophysics, School of Physics and Astronomy, Monash University, VIC 3800, Australia\\
$^{3}$Observatoire de l'Universit\'e de Gen\`eve, 51 chemin des Maillettes, 1290 Versoix, Switzerland\\
$^{4}$Department of Astronomy $\&$ Astrophysics, 5640 S. Ellis Ave, Chicago, IL 60637, USA\\
$^{5}$Center for Computational Astrophysics, Flatiron Institute, 162 5th Ave, New York, NY 10010, USA\\
$^{6}$Research School of Astronomy and Astrophysics, The Australian National University, Cotter Road, Canberra, ACT 2611, Australia\\
$^{7}$Tacoma Community College, 6501 South 19th Street, Tacoma, WA 98466-7400, USA\\
$^{8}$Universidade Federal do Rio Grande do Sul, Instituto de F\'isica, Av. Bento Gon\c{c}alves 9500, Porto Alegre, RS, Brazil\\
$^{9}$Institut f{\"u}r Astrophysik, Georg-August Universit{\"a}t G{\"o}ttingen, Wilhelmsplatz 1, 37073, G{\"o}ttingen, Germany
}

\date{Accepted 10/11/2017. Received 30/9/2017.}

\pubyear{2017}

\begin{document}
\label{firstpage}
\pagerange{\pageref{firstpage}--\pageref{lastpage}}
\maketitle

\begin{abstract}
Important insights into the formation and evolution of the Galactic disc(s) are contained in the chemical compositions of stars. We analysed high-resolution and high signal to noise HARPS spectra of 79 solar twin stars in order to obtain precise determinations of their atmospheric parameters, ages ($\sigma$$\sim$0.4 Gyr) and chemical abundances ($\sigma$$<$0.01~dex) of 12 neutron-capture elements (Sr, Y, Zr, Ba, La, Ce, Pr, Nd, Sm, Eu, Gd, and Dy). This valuable dataset allows us to study the [X/Fe]-age relations over a time interval of $\sim$10 Gyr and among stars belonging to the thin and thick discs. These relations show that i) the $s$-process has been the main channel of nucleosynthesis of $n$-capture elements during the evolution of the thin disc; ii) the thick disc is rich in $r$-process elements which suggests that its formation has been rapid and intensive. %; iii) a chemical continuity between the thin and thick discs is evident in the abundances of Ba. 
In addition, the heavy (Ba, La, Ce) and light (Sr, Y, Zr) $s$-process elements revealed details on the dependence between the yields of AGB stars and the stellar mass or metallicity. Finally, we confirmed that both [Y/Mg] and [Y/Al] ratios can be employed as stellar clocks, allowing ages of solar twin stars to be estimated with an average precision of $\sim$0.5~Gyr.
\end{abstract}


%%%%%%%%%%%%%%%%%%%%%%%%%%%%%%%%%%%%%%%%%%%%%%%%%%

%%%%%%%%%%%%%%%%% BODY OF PAPER %%%%%%%%%%%%%%%%%%

\section{Introduction}
\label{intro}

The history of matter in galaxies is written in the chemical composition of stellar populations. Akin to archaeologists who use fossils to infer the past of ancient civilisations, astronomers can decrypt the information locked into stellar spectra to trace the history of the Milky Way.

The pristine material from which our Galaxy has formed was extremely poor in metals. All the elements heavier than Li have been subsequently produced in stars through different sites of nucleosynthesis (e.g., Type II or Type Ia supernovae, asymptotic giant branch stars), with their own unique pattern of species produced and delivered into the insterstellar medium (ISM). Due to this variety of channels, the abundance ratios of two species having different origin change with time and can be used to study the formation and evolution of stellar populations (e.g., \citealp{Gilmore89,Chiappini97,RecioBlanco14,RojasArriagada16,RojasArriagada17}). Recent studies have shown that the knowledge of the relations between the abundance ratios and the stellar ages can provide insights on the variables that are controlling the evolution of our Galaxy, such as the star formation rate, the initial mass function, the mass and metallicity of the supernovae (SNe) progenitors, and their yields \citep{Haywood15,Snaith15,Battistini16,Nissen15,Nissen16,Spina16,Spina16b}. 

These latter studies proceed along a well defined path enabled by an increasingly precise abundance analysis of stars that aims to unravel the chemical enrichment history of the Galactic disc. In this context, the study of the elements heavier than zinc (Z>30) play a crucial role, as they trace yields of a broad range of sites of nucleosynthesis with very different timescales \citep{Sneden08}. These species are produced via neutron capture of isotopes heavier than iron: neutrons are captured by nuclei that, if unstable, can $\beta$ decay transforming neutrons into protons. The group of elements that are synthesised through this process are commonly dubbed as neutron- ($n$)-capture elements. The $n$-capture processes can occur through two channels: if the neutron capture timescale is longer than the decay timescale, it is called the slow- ($s$-) process, otherwise a rapid- ($r$-) process takes place. The $n$-capture timescales are largely determined by the density of free neutrons (n$_{n}$): the $s$-process requires n$_{n}$$\lesssim$10$^{8}$ cm$^{-3}$ \citep{Busso99}, while the $r$-process occurs with much higher densities (n$_{n}$$\sim$10$^{24-28}$ cm$^{-3}$; \citealt{Kratz07}). This also indicates that the two channels require very different astrophysical environments.

The $s$-process is believed to occur in low- and intermediate-mass stars (i.e., 0.5-8 M$_{\odot}$) during their asymptotic giant branch (AGB) phase, when they experience recurrent pulses driven by cyclical thermal instabilities in the He burning shell \citep{Gallino98,Sneden08}. The products of this burning are mixed into the stellar surface by periodic convective mixing episodes, known as dredge-up events (see \citealt{Karakas14}). Strong stellar winds then expel this enriched material into the ISM contributing to the Galactic chemical evolution. 
During the $s$-process, the synthesis of new elements 
%During the $s$-process, the free neutrons produced through different mechanisms can be captured by Fe or heavier elements synthesising new species. This process 
moves through a path of stable nuclei until it encounters three bottlenecks corresponding to the elements with number of neutrons N= 50, 82, and 126, which are more stable against neutron capture than other nuclei. Therefore, the flux of neutron captures tends to accumulate nuclei at these barriers until it reaches a large enough value to make the probability of $n$-captures on these species significant. Once the bottleneck is bypassed, the $s$-process path proceeds to the next bottleneck. These congestions in the path shape the final distribution of $s$-process products, which is characterised by three main peaks. The first peak at N=50 is formed by the so-called $\textit{light}$ $s$-process elements ($ls$; Sr, Y, and Zr). The second peak at N=82 comprises the $\textit{heavy}$ $s$-process elements ($hs$; Ba, La, and Ce). The third peak at N=126 (Pb) is the termination of the $s$-process path (see \citealt{Lugaro12}). 

There are two sources of free neutrons in AGB stars. The primary source is the $^{13}$C which is activated in low-mass stars, while the $^{22}$Ne neutron source is activated in higher-mass stars \citep{Gallino98,Busso01}. At solar metallicities, the $^{13}$C neutron source is dominant for AGB stars of 1.5-3 M$_{\odot}$, while the $^{22}$Ne source is prominent for masses $\gtrsim$5 M$_{\odot}$ \citep{Goriely04}. In stars with masses in-between the two regimes, a combination of $^{13}$C and $^{22}$Ne neutron sources is in action. The production of $s$-process elements is strongly dependent on the initial stellar mass (e.g., see AGB models by \citealt{Karakas16}). Namely, since the $^{13}$C neutron source is the responsible for the production of the bulk of the $s$-process elements, the nucleosynthesis of these species peaks for low-mass AGB stars when the $^{13}$C is the dominant source. In this regime, the high flux of free neutrons permits a great production of $hs$ elements at expenses of the $ls$ elements. %On the other hand, the $^{22}$Ne source produces a smaller flux of free neutrons and, consequently, less $s$-process elements are synthesised. As a result, at higher masses the production of $hs$ and $ls$ elements is equally negligible.
On the other hand, the $^{22}$Ne source operates over much shorter time scales than the $^{13}$C source and consequently fewer $s$-process elements are synthesised. As a result, at higher masses the production of $ls$ and $hs$ elements is equally negligible in models around solar metallicity.

%Despite the high uncertainties and assumptions that affect the AGB models, \citet{Karakas16} have calculated the yields of $s$-process elements weighted by the Salpeter initial mass function and have shown that the production of these elements is strongly dependent on the initial stellar mass and metallicity (see also the models by \citealt{Cristallo15,Pignatari16}). Namely, since the $^{13}$C neutron source is the responsible for the production of the bulk of the $s$-process elements, the nucleosynthesis of these species peaks for low-mass AGB stars when the $^{13}$C is the dominant source (i.e., at solar metallicities, in a mass range between 1.5-3 M$_{\odot}$). In this regime, the high flux of free neutrons permits a great production of hs elements at expenses of the ls elements. On the other hand, the $^{22}$Ne source produces a smaller flux of free neutrons and, consequently, less $s$-process elements are synthesised, with a bias toward the ls elements. As a result, at higher masses the relative production of hs and ls elements tends to equalise. 

The s-process production is also strongly dependent on the stellar metallicity \citep{Cristallo15,Pignatari16,Karakas16}. At solar and super-solar metallicities, the production of s-process elements relative to Fe, [s/Fe], is negligible when the $^{22}$Ne is the dominant source of free neutrons, but the models predict an increment of the [s/Fe] ratios at lower metallicities (i.e., Z$\lesssim$1/2Z$_{\odot}$) over the entire range of AGB stellar masses. The low-mass stars, with the $^{13}$C as the dominant neutron source, are those that contribute the most to the production of $s$-process elements regardless of the initial stellar metallicity. Therefore, since the low-mass stars have longer lifetimes than those of intermediate-mass, the greatest contribution to the chemical evolution of the Milky Way from AGB stars is expected to have occurred - and it is still occurring - during the latest evolution of the Galaxy, when the thin disc was already formed \citep{Busso01}. In addition, the AGB models also predict that, at a given stellar mass, the [$hs$/$ls$] ratio yielded by progenitors with sub-solar metallicity is higher than that of AGB stars with solar metallicity and that the production of $light$ $s$-process elements increases with stellar metallicity until it becomes nearly equal to that of $heavy$ $s$-process which occurs, for low-mass stars, at Z$\gtrsim$2Z$_{\odot}$ \citep{Karakas16}.

The $r$-process likely occurs in the neutrino-driven winds of core-collapse SNe \citep{Woosley95,Wanajo13}, or during the merging of two neutron stars or a neutron star and a black hole \citep{Argast04,Surman08,Korobkin12}. 
Even if the actual production site(s) of the $r$-process is not very well known at present \citep{Cowan04,Thielemann11}, recently \citet{Rosswog14} has shown that the neutrino-driven winds can be the main producers of the light $r$-process elements (those with A from 50 to 130), while the mergers of compact objects would mainly synthesise the heavy $r$-process elements (those with A$\geq$130). %This would imply that the oldest components of our Galaxy (e.g., halo, thick disc) should be richer of light $r$-process elements than the thin disc stars.

Most $n$-capture elements can be produced through both $s$- and $r$-processes. Therefore, it is not always straightforward to disentangle the channels and the progenitors that are involved in the evolutionary path of each element. Recently, \citet{Bisterzo14} were able to predict the production rates for the $n$-capture elements at the epoch of the solar system formation. Their study has shown that the $s$-process is responsible for 85$\%$ of the solar Ba, which thus can be considered a prototype of the $s$-process elements. On the other hand, Eu is almost a pure $r$-process element, since only 6$\%$ of its abundance is produced through the $s$-process.

In the present paper we aim to provide new insights on the [X/Fe] vs age relations for 12 $n$-capture elements from Sr to Dy, through the exploitation of high-quality data of solar twins spanning ages from 0.5 to 10 Gyr. For some species, these correlations have previously been investigated by \citet{Nissen15}, \citet{Spina16}, and \citet{Spina16b}, through samples of 21, 13, and 41 solar twin stars, respectively. In the present work, we are extending these studies to a larger number of elements and using a sample of 79 solar twin stars. In Section 2 we describe the spectral analysis and the determination of the chemical abundances and stellar ages. The distributions of the abundance ratios as a function of the stellar ages are discussed in Section 3. In Section 4 we present our concluding remarks. In a forthcoming paper, Bedell et al. (in preparation), we will employ the same dataset to study the [X/Fe]-age distributions of the lighter elements (i.e., those from C to Zn).

%%%%%%%%%%%%%%%%%%%%%%%%%%%%%%%%%%%%%%%%%%%%%%%%%%
%%%%%%%%%%%%%%%%%%%%%%%%%%%%%%%%%%%%%%%%%%%%%%%%%%
\begin{table*}
	\centering
	\caption{The list of the 79 solar twins stars analysed in this paper.}
	\label{literature}
	\begin{threeparttable}
		\begin{tabular}{llllll}
			\hline
			HIP & HD & V & $\pi$ & Ref. & Binarity \\
			 &  & (mag) & (mas) & & \\ \hline
			10175 & 13357 & 8.180 $\pm$ 0.016 & 23.32 $\pm$ 0.28 & Gaia & \\
			101905 & 196390 & 7.328 $\pm$ 0.008 & 29.47 $\pm$ 0.59 & Gaia & \\
			102040 & 197076 & 6.425 $\pm$ 0.004 & 47.79 $\pm$ 0.75 & Hipparcos & \\
			102152 & 197027 & 9.156 $\pm$ 0.024 & 13.08 $\pm$ 0.97 & Gaia & \\
			10303 & 13612B & 7.629 $\pm$ 0.013 & 23.43 $\pm$ 9.50 & Hipparcos & \\
			... & ... & ... & ... & ... & ... \\ \hline
			\end{tabular}
		\begin{tablenotes}
			\item $\bf{Note.}$ The full version of this table is available online at the CDS.
		\end{tablenotes}
	\end{threeparttable}
\end{table*}

\section{Spectroscopic analysis}
\label{Spectroscopicanalysis}
Here we describe the sample of 79 solar twins used for this study and discuss the reduction and analysis of the HARPS spectra. As detailed below, high-precison determinations of the stellar parameters, ages and chemical abundances are the main products of our analysis.

\subsection{Spectroscopic sample and data reduction}
The 79 stars analysed in this paper have been selected from the sample presented in \citet{Ramirez14} and are objects with atmospheric parameters that are very similar to those of the Sun, within $\pm$100 K in T$_{\rm eff}$ and $\pm$0.1 dex in log g and [Fe/H]. The stars with parameters that match these criteria are called ``solar twins''. As it has been shown by several authors (e.g., \citealt{Bedell14, Biazzo15, Nissen15,YanaGalarza16a,Spina16,Spina16b}), a differential analysis relative to the Sun permits to obtain chemical abundances at sub-0.01~dex precision and age uncertainties of $\sim$0.5~Gyr for solar twins. 

These stars have been observed with the HARPS spectrograph \citep{Mayor03} on the 3.6 m telescope at the La Silla observatory, mostly through our ESO Large Programme (ID: 188.C-0265) that aimed to search for planetary systems around solar twin stars (e.g., \citealt{Bedell15,Melendez17}). We have also employed data from other programmes\footnote{In addition to the observations collected by the ESO Programme 188.C-0265, we have also used the spectra of the programmes 183.D-0729, 292.C-5004, 077.C-0364, 072.C-0488, 092.C-0721, 093.C-0409, 183.C-0972, 192.C-0852, 091.C-0936, 089.C-0732, 091.C-0034, 076.C-0155, 185.D-0056, 074.C-0364, 075.C-0332, 089.C-0415, 60.A-9036, 075.C-0202, 192.C-0224, 090.C-0421 and 088.C-0323.} in order to increase the data set. In addition to the solar twins, the sample includes a number of solar spectra acquired through HARPS observations of the asteroid Vesta.

In a recent paper \citet{dosSantos17} has studied the radial velocity modulations of the objects in our sample, identifying 15 spectroscopic single-lined binary systems. An additional star, HIP 77052, has been found by \citet{Tokovinin14} to be part of a very tight visual binary system with angular separation $\sim$5''. In Table \ref{literature} we list the stars included in our sample, together with other relevant information: the V magnitudes \citep{Kharchenko09}, the parallaxes from Gaia \citep{Gaia16} or Hipparcos \citep{Kharchenko09}, and the information on binarity \citep{dosSantos17,Tokovinin14}.% This latter also reports the angular separations between the components of visual binaries: we used this information, together with the parallaxes to calculate the physical separations listed in the last column of Table~\ref{literature}.

\begin{table*}
\centering
\caption{Atmospheric parameters determined through our analysis.}
\begin{threeparttable}
\label{parameters}
\begin{tabular}{cccccccc}
\hline
Star & S/N & T$_{\rm eff}$   & log g   & [Fe/H]   & $\xi$   \\ 
 & (pxl$^{-1}$) & (K)  & (dex) & (dex) & (km s$^{-1}$) \\ \hline
HIP10175 & 600 & 5719 $\pm$ 3 & 4.485 $\pm$ 0.010 & -0.028 $\pm$ 0.002 & 0.97 $\pm$ 0.01 \\
HIP101905 & 1200 & 5906 $\pm$ 5 & 4.500 $\pm$ 0.011 & 0.088 $\pm$ 0.004 & 1.08 $\pm$ 0.01 \\
HIP102040 & 1000 & 5853 $\pm$ 4 & 4.480 $\pm$ 0.012 & -0.080 $\pm$ 0.003 & 1.05 $\pm$ 0.01 \\
HIP102152 & 600 & 5718 $\pm$ 4 & 4.325 $\pm$ 0.011 & -0.016 $\pm$ 0.003 & 0.99 $\pm$ 0.01 \\
HIP10303 & 700 & 5712 $\pm$ 3 & 4.395 $\pm$ 0.010 & 0.104 $\pm$ 0.003 & 0.94 $\pm$ 0.01 \\
... & ... & ... & ... & ... & ... \\ 
\hline
\end{tabular}
\begin{tablenotes}
\item $\bf{Note.}$ The full version of this table is available online at the CDS.
\end{tablenotes}
\end{threeparttable}
\end{table*}

The wavelength coverage of the HARPS spectrograph is 3780 to 6910 $\AA$, with a spectral resolving power of R = 115~000. Data reduction was performed automatically with the HARPS Data Reduction Software, which also calculates the spectral Doppler shift for each exposure. We used IRAF's \texttt{dopcor} task to correct each exposure for the Doppler shift, the 
\texttt{scombine} task to merge all the exposures and create a single-column FITS spectrum for each star and the \texttt{continuum} task to normalise the final spectra.  The final signal-to-noise ratios (S/N) are within 300 and 1800 pixel$^{-1}$ at 600 nm with a median of 800 pixel$^{-1}$. The normalised spectra have been used for the spectroscopic analysis.

\begin{figure*}
\centering
	\includegraphics[width=17.5cm]{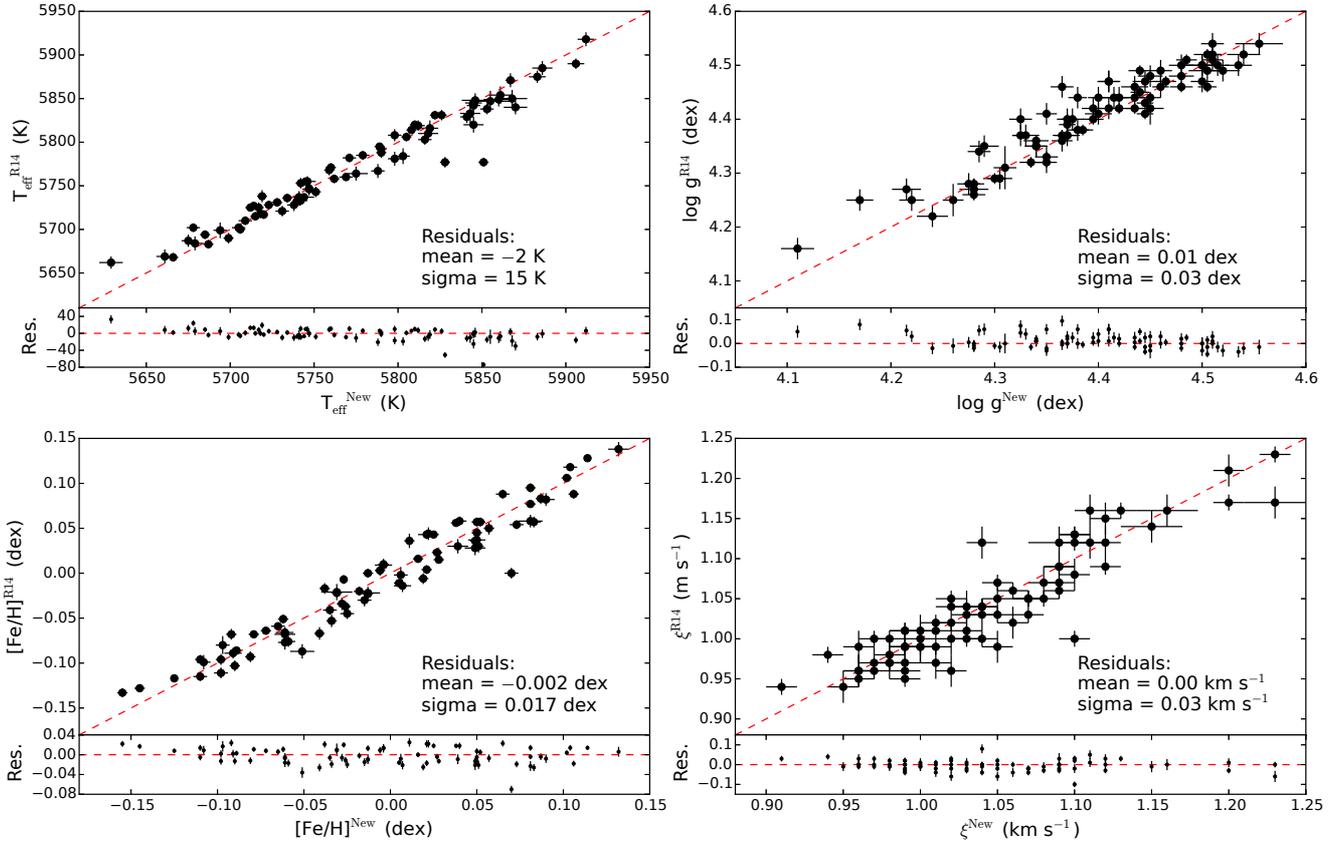}
    \caption{In each panel we compare the stellar parameters (T$_{\rm eff}$, log~g, [Fe/H], and $\xi$) determined by \citet{Ramirez14b} (ordinate) to those obtained through our analysis (abscissa). For each parameter we show, in the upper panels, the values with their uncertainties and, in the lower panels, the residuals. The red dashed lines indicate the locus of residuals equal to zero.}
    \label{comparisons_params}
\end{figure*}

\subsection{Stellar parameters and abundances}
Our method of analysis of the solar twin spectra is similar to that performed by \citet{Spina16,Spina16b}. Namely, we used the master list of atomic transitions employed by \citet{Melendez14} that includes 98 lines of Fe I, 17 of Fe II, and 35 lines of 12 $n$-capture elements (i.e., Sr, Y, Zr, Ba, La, Ce, Pr, Nd, Sm, Eu, Gd, and Dy) detectable in the HARPS spectral range. Through the IRAF \texttt{splot} task and adopting the same approach described in \citet{Bedell14}, we measured the equivalent widths (EWs) of these absorption features in all the spectra.

The EW measurements are processed by the \textit{qoyllur-quipu}\footnote{Qoyllur-quipu or q$^2$ \citep{Ramirez14b} is a Python package that is free and available online at \url{https://github.com/astroChasqui/q2}.} code that automatically estimates the stellar parameters by performing a line-by-line differential excitation/ionisation balance analysis of the iron EWs relative to the solar spectrum. Specifically, the q$^2$ code iteratively searches for the three equilibria (excitation, ionisation, and the trend between the iron abundances and the reduced equivalent width log[EW/$\lambda$]). The iterations are executed with a series of steps starting from a set of initial parameters (i.e., the nominal solar parameters) and arriving at the final set of parameters that simultaneously fulfil the equilibria. Further details on this procedure can be found in \citet{Ramirez14b}. For this analysis we employed the Kurucz (ATLAS9) grid of model atmospheres \citep{Castelli04} and we assumed the following solar parameters: T$_{\rm eff}$=5777~K, log g=4.44 dex, [Fe/H]=0.00 dex and $\xi$ =1.00 km~s$^{-1}$ (e.g., \citealt{Cox00}). The errors associated with the stellar parameters are evaluated by the code following the procedure described in \citet{Epstein10} and \citet{Bensby14}, that takes into account the dependence between the parameters in the fulfilment of the three equilibrium conditions. The typical precisions that we reached are $\sigma$(T$_{\rm eff}$)=4 K, $\sigma$(log g)=0.012 dex, $\sigma$([Fe/H])=0.004, $\sigma$($\xi$)=0.011 km/s. The stellar parameters and their uncertainties are reported in Table \ref{parameters}.

The atmospheric parameters of the stars in our sample
%, with the exception of HIP~114328,
 have been also determined by \citet{Ramirez14b}. They used the line-by-line differential technique on a sample of MIKE spectra with a resolving power of 83~000 (65~000) in the blue (red) CCD and a typical S/N of 400. In Fig. \ref{comparisons_params} we compare the results obtained by Ramirez et al. (R14) with those from our analysis. Considering the mean of the residuals, there is a general agreement between the two datasets. However, the stellar parameters of a large fraction of the stars are not consistent within the errors. These discrepancies could be attributed only in part to the different linelist and the atmospheric models employed by R14 or to the lower S/N of the MIKE spectra. In fact, through a visual inspection of the MIKE spectra analysed by R14, we have noticed that they are affected by some complications in the wavelength calibration across the blue chip. Also, the spectral resolution of the MIKE spectrograph is not stable, and is lower than in the HARPS spectrograph. These issues and the lower quality of the MIKE spectra, are likely the main causes of the scatter observed in Fig. \ref{comparisons_params}. 

Once the stellar parameters and the relative uncertainties have been determined for each star, q$^2$ automatically employs the appropriate atmospheric model for the local thermodynamic equilibrium (LTE) calculation of the chemical abundances through the 2014 version of MOOG \citep{Sneden73}. All the elemental abundances are scaled relatively to the values obtained for the Sun on a line-by-line basis. 
In addition, through the \texttt{blends} driver in the MOOG code and adopting the line list from the Kurucz database, q$^2$ was able to take into account the hyperfine splitting effects in the abundance calculations of Y, Ba, La, Pr and Eu. Finally, the q$^2$ code determined the error budget associated with the abundances [X/H] by summing in quadrature the observational error due to the line-to-line scatter from the EW measurements (standard error), and the errors in the atmospheric parameters. When, as for Sr, Pr, Eu, and Gd, only one line is detected, the observational error was estimated by repeating the EW measurement five times with different assumptions on the continuum setting\footnote{As it is detailed in \citet{Bedell14}, an accurate measurement of equivalent width would depend on finding the true spectral continuum. In order to minimise the impact of nearby features, our approach employed the point(s) in the immediate vicinity of the line (i.e., about $\pm$3$m\AA$) that appears almost constant across the HARPS spectra of our sample. However, different point(s) could be used to estimate the local continuum of each line.}, adopting as error the standard deviation. The chemical abundances of the $n$-capture elements are listed in Table. \ref{abundance}.

\subsection{Stellar ages and masses}
\label{ages_masses}

It is well known that a star, as it ages, evolves along a determined track in the Hertzsprung-Russell diagram that - to a first approximation - depends on the stellar mass and metallicity. Therefore, if the atmospheric parameters and the absolute magnitude M$_{\rm V}$ of the star are known with enough precision, it will be possible to determine reasonable estimates of its age and mass. Through this approach, that is commonly dubbed as the \textit{isochrone method} (e.g., \citealt{Vandenberg85,Lachaume99}), the q$^2$ code allows us to calculate the age and mass probability distributions for each star of our sample, by employing a grid of isochrones and the set of stellar parameters determined through our analysis along with their uncertainties. Details on the exact procedure followed by q$^2$ are given in \citet{Ramirez14,Ramirez14b}. In short, given a grid of isochrones, the code calculates the probability distribution $p$ of the stellar ages and masses using as weights the differences between the observed T$_{\rm eff}$, log~g, and [Fe/H] values for a star (normalised by their errors) and the corresponding values in the isochrones' grid (T, log G, and [M/H]): 

\begin{equation}
\begin{split}
p \propto &~ {\rm exp[-(T_{\rm eff}-T)^2/2(\delta T_{\rm eff})^2)]} \\
 & \times{\rm exp[-(log g- log G)^2/2(\delta log g)^2)]} \\
 & \times{\rm exp[-([Fe/H]-[M/H])^2/2(\delta [Fe/H])^2)]}.
\end{split}
\end{equation}

By default the q$^2$ code performs a maximum-likelihood calculation to determine which age and mass values are the most probable (i.e., the peak of the probability distribution). It also calculates the 68$\%$ (1-$\sigma$-like) and 95$\%$ (2-$\sigma$-like) credible intervals for these parameters, as well as the simple mean and standard deviation values of all isochrone points that fall inside the T$_{\rm eff}$ - log~g - [Fe/H] space covered by the uncertainties on these parameters. 
Hereafter, the stellar age values that we will use are those corresponding to the main peaks of the probability distributions and their uncertainties coincide with the 1$\sigma$ credible intervals.

For the calculation we used the grid of Yonsei-Yale isochrones \citep{Yi01,Kim02} shifted by +0.04 dex in metallicity [M/H], that is necessary to take into account the influence of atomic diffusion at the solar age \citep{Melendez12,Dotter17}. Through this normalisation and assuming the standard solar parameters, we were able to recover a solar age of 4.6 Gyr, as expected for the age of the solar system \citep{Connelly08,Amelin10}. We also performed an additional adjustments of the isochrones' metallicity to simulate the effects of $\alpha$-enhancement on the model atmospheres. 
For this further shift we have employed Eq. 3 in \citet{Salaris93}:

\begin{equation}
{\rm[M/H]=[Fe/H]}+\log(0.638\times10^{\rm[\alpha/Fe]}+0.362),
\end{equation}

where we used the stellar Mg abundances, determined by Bedell et al. (in prep), as a proxy of the $\alpha$-abundances. 
%The age and mass values produced through this calculation are listed in Table \ref{ages_logg}, where Age$_{\rm mp}$ is the age corresponding to the peak of the probability distribution, Age$_{\rm ll-1\sigma}$ and Age$_{\rm ul-1\sigma}$ are the lower and upper limits of the 68$\%$ confidence interval in the probability distribution, Age$_{\rm ll-2\sigma}$ and Age$_{\rm ul-2\sigma}$ are the lower and upper limits of the 95$\%$ confidence interval in the probability distribution, and Age$_{\rm mean}$ is the simple mean of all isochrone points that fall inside the space of parameters employed for this calculation. The same notation is used for the mass values. Hereafter, the stellar age values that we will use are those corresponding to the main peaks of the probability distributions and their uncertainties will coincide with the 1$\sigma$ confidence intervals. Three young stars (i.e., HIP 3203, HIP 4909, and HIP 5412) have the probability age distributions that are truncated before than they reach a maximum, in these cases we will adopt the Age$_{\rm ul-1\sigma}$ as upper limit of their ages.

\begin{figure}
\centering
	\includegraphics[width=8.5cm]{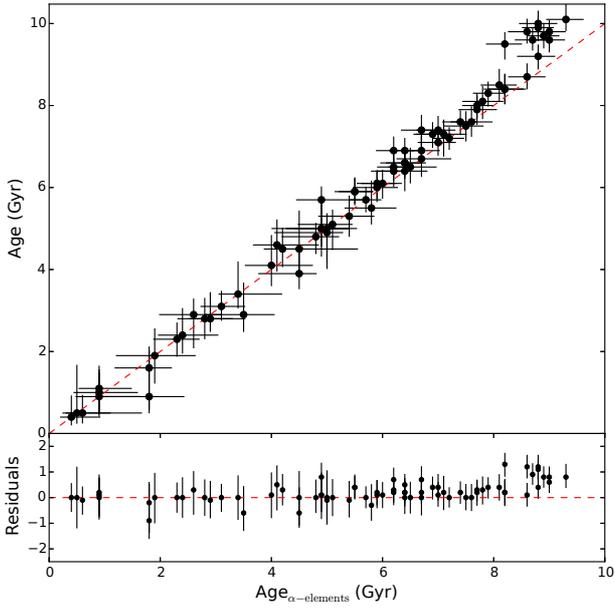}
    \caption{Comparison of stellar ages derived including the $\alpha$-enhancement (abscissa) with those derived by neglecting this effect (ordinate). The red dashed line indicates the locus of residuals equal to zero.}
    \label{comparison_alpha}
\end{figure}

In Fig. \ref{comparison_alpha} we explore the impact of having considered in our calculations the $\alpha$-abundance in the stellar atmospheres: the ages agree very well, with a small ($\sim$1~Gyr) systematic deviation for the oldest stars, those with the highest [$\alpha$/Fe] ratios (e.g., [Mg/Fe]$\gtrsim$0.10~dex; \citealt{Spina16b}). A similar result was found by \citet{Nissen15}.

\begin{figure}
\centering
	\includegraphics[width=8.5cm]{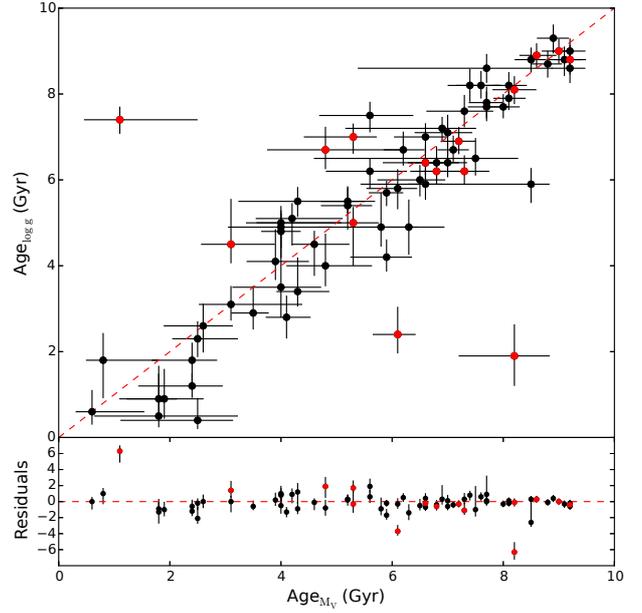}
    \caption{Comparison between the ages obtained through the stellar log~g (age$_{\rm log~g}$) with those determined with the M$_{\rm V}$ (age$_{\rm Mv}$). The red dots indicate the spectroscopic binaries and the star HIP~77052, member of a tight visual binary system. The red dashed line indicates the locus of residuals equal to zero.}
    \label{comparison_logg_plx}
\end{figure}

The q$^2$ code also allows to employ the absolute magnitude M$_{\rm V}$ instead of log~g. Therefore, by determining the M$_{\rm V}$ values calculated from the V magnitudes and the parallaxes\footnote{Interstellar reddening is important only for stars more distant than about 100 pc. As shown by \citet{Lallement14}, the Sun is located at the centre of a 200 pc-wide cavity, called Local Bubble, which appears to be dust free and is characterised by dE(B-V)/dr$\leq$0.0002~mag pc$^{-1}$. Since all the stars in our sample are located within this volume and since this reddening is negligible in comparison with the typical errors in parallaxes, we have assumed a E(B-V)=0.} listed in Table~\ref{literature}, we have repeated the calculation through this alternative procedure. 
%The q$^2$ outputs are reported in Table~\ref{ages_Mv}. 
In Fig. \ref{comparison_logg_plx} we compare the mean values of the ages obtained through the stellar log~g (i.e., age$_{\rm log~g}$) with those determined with the M$_{\rm V}$ (i.e., age$_{\rm Mv}$). 
As expected, there is an excellent agreement between the two procedures for the age estimation, with the exception of some binary stars that have highly discrepant age determinations. It is likely that the V magnitudes of these objects include a contribution from the companions' flux, which affected the determinations of age$_{\rm Mv}$. %In addition, among the 63 single stars, there are 11 objects whose residuals are marginally inconsistent with zero within the 1-$\sigma$ error.

In view of the general and satisfactory agreement between age$_{\rm log~g}$ and age$_{\rm Mv}$, we have modified the q$^2$ code in order to exploit, at the same time, the log~g and M$_{\rm V}$ values for the calculation of the age and mass probability distributions. The new probability distributions are determined as follows:

\begin{equation}
\begin{split}
p[{\rm log\ g\ \wedge\ M_{V}}] \propto &~ {\rm exp[-(T_{\rm eff}-T)^2/2(\delta T_{\rm eff})^2)]} \\
 & \times{\rm exp[-(log~g-log G)^2/2(\delta log~g)^2)]} \\
 & \times{\rm exp[-([Fe/H]-[M/H])^2/2(\delta [Fe/H])^2)]} \\
 & \times{\rm exp[-(M_{\rm v}-V)^2/2(\delta M_{\rm v})^2)]}.
\end{split}
\label{equation}
\end{equation}

\begin{figure*}
\centering
	\includegraphics[width=17.5cm]{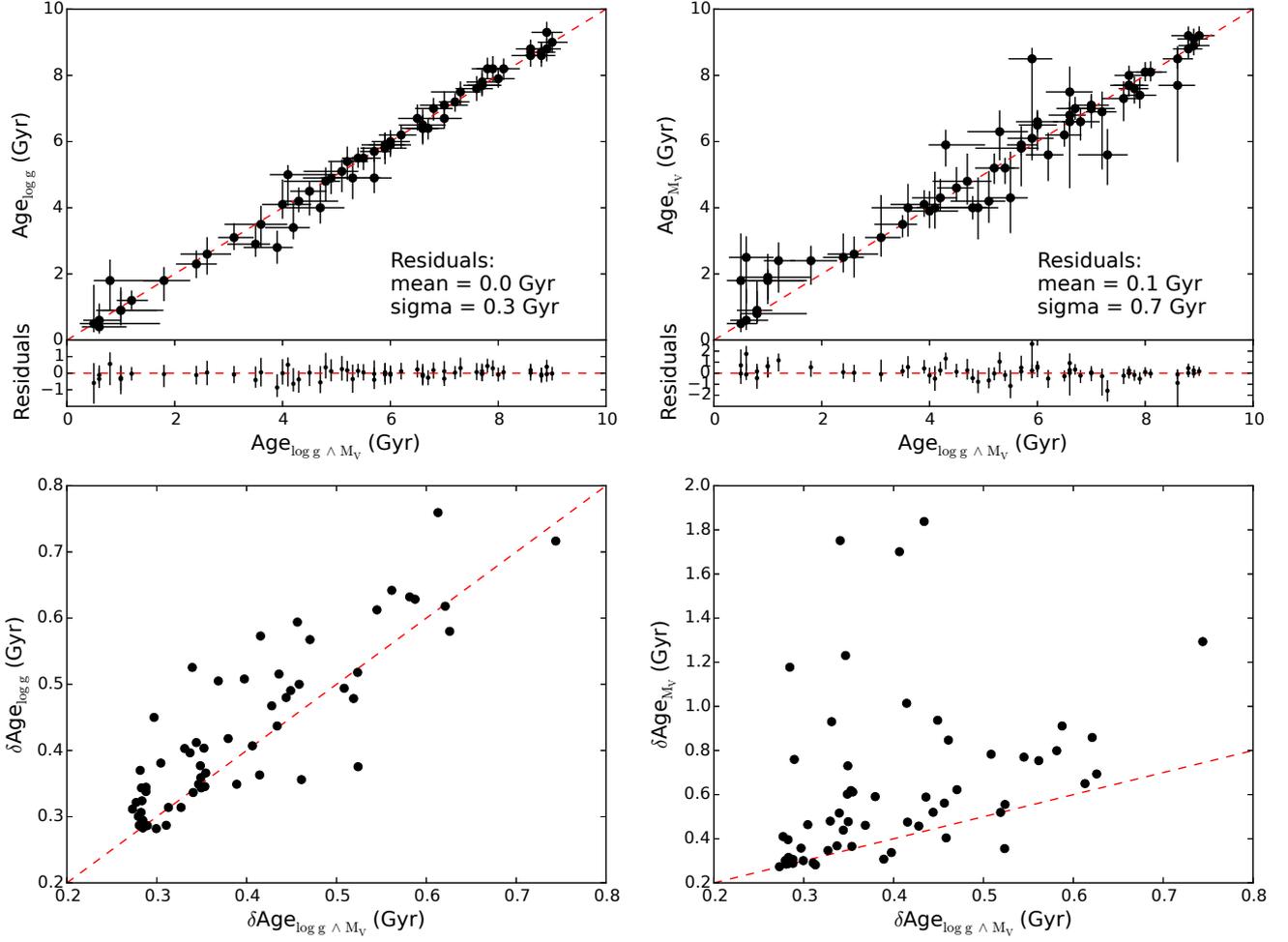}
    \caption{\textbf{Top panels.} Comparison between age determinations obtained through Equation \ref{equation} (i.e., age$_{\rm log g \wedge Mv}$) with the age values determined through the standard procedures of q$^2$ (i.e., age$_{\rm log~g}$ and age$_{\rm Mv}$). The red dashed line indicates the locus of equal values. \textbf{Bottom panels.} Comparison between the age uncertainties (i.e., half widths of the 68$\%$ credible intervals)obtained through the different approaches.}
    \label{comparisons_ages}
\end{figure*}

We have employed Eq.~\ref{equation} to make a third calculation of the ages for all the 63 single stars in our sample. In the two panels on the top of Fig. \ref{comparisons_ages} we compare these new age determinations (age$_{\rm log g \wedge Mv}$) with the age values determined through the standard procedures of q$^2$ (age$_{\rm log~g}$ and age$_{\rm Mv}$), while in the two bottom panels we compare the uncertainties obtained through the different approaches. As expected, there is a remarkable agreement between the three procedures of age determinations. However, comparison between age$_{\rm log~g}$ and age$_{\rm log g \wedge Mv}$ is that with a greater agreement (i.e., $\sigma$=0.3~Gyr), indicating that the spectroscopic log~g values are still a superior luminosity indicator for bright solar twins than the parallaxes from Gaia DR1 or Hipparcos plus the V magnitudes. 
%the precision of our spectroscopic log~g values permits to put more stringent constraints on the age determinations than the parallaxes from Gaia DR1 or Hipparcos plus the V magnitudes. 
In addition, as shown in the bottom panels of Fig.~\ref{comparisons_ages}, the age uncertainties obtained through our new approach described in Eq.~\ref{equation} are typically smaller than those produced by classical approaches of q$^2$. In fact, the simultaneous use of the independent observables M$_{\rm V}$ and log~g %in the interpolation with the grid of isochrones 
permits to reduce the isochrone points that fit the space of parameters covered by each star. This leads to a general increase of precision in the age and mass determinations. 

Hereafter, we adopt as our final age values the age$_{\rm log g \wedge Mv}$ determinations for all the single stars, while for the spectroscopic binaries and for HIP~77052 we use the age$_{\rm log~g}$ determinations. The final age estimates are listed in Table~\ref{ages_Mv_logg}, where Age$_{\rm mp}$ is the age corresponding to the peak of the probability distribution, Age$_{\rm ll-1\sigma}$ and Age$_{\rm ul-1\sigma}$ are the lower and upper limits of the 68$\%$ credible interval in the probability distribution, Age$_{\rm ll-2\sigma}$ and Age$_{\rm ul-2\sigma}$ are the lower and upper limits of the 95$\%$ credible interval in the probability distribution, and Age$_{\rm mean}$ is the simple mean of all isochrone points that fall inside the space identified by the stellar parameters and their uncertainties employed for this calculation. The same notation is used for the mass values. Two young stars (i.e., HIP 3203, and HIP 4909) have the probability age distributions that are truncated before than they reach a maximum, in these cases we will adopt the Age$_{\rm ul-1\sigma}$ as upper limit of their ages. The typical uncertainties of our age determinations (i.e., the average of the half widths of the 68$\%$ credible intervals) is equal to 0.4 Gyr.

\begin{landscape}
\begin{table}
\centering
\tiny
\caption{Chemical abundances of the $n$-capture elements determined for the sample of 79 solar twin stars.}
\begin{threeparttable}
\label{abundance}
\begin{tabular}{ccccccccccccc}
\hline
Star & [Sr I/H] & [Y II/H] & [Zr II/H] & [Ba II/H] & [La II/H] & [Ce II/H] & [Pr II/H] & [Nd II/H] & [Sm II/H] & [Eu II/H] & [Gd II/H] & [Dy II/H] \\ \hline
HIP10175 & 0.053 $\pm$ 0.005 & 0.038 $\pm$ 0.007 & 0.059 $\pm$ 0.012 & 0.088 $\pm$ 0.005 & 0.09 $\pm$ 0.029 & 0.076 $\pm$ 0.016 & 0.142 $\pm$ 0.008 & 0.099 $\pm$ 0.01 & 0.029 $\pm$ 0.013 & 0.061 $\pm$ 0.005 & 0.052 $\pm$ 0.005 & 0.037 $\pm$ 0.005 \\
HIP101905 & 0.185 $\pm$ 0.006 & 0.169 $\pm$ 0.011 & 0.173 $\pm$ 0.01 & 0.228 $\pm$ 0.01 & 0.19 $\pm$ 0.031 & 0.207 $\pm$ 0.012 & 0.182 $\pm$ 0.014 & 0.219 $\pm$ 0.01 & 0.142 $\pm$ 0.021 & 0.164 $\pm$ 0.014 & 0.137 $\pm$ 0.006 & 0.154 $\pm$ 0.066 \\
HIP102040 & -0.018 $\pm$ 0.005 & -0.026 $\pm$ 0.007 & 0.001 $\pm$ 0.011 & 0.048 $\pm$ 0.012 & 0.038 $\pm$ 0.018 & 0.069 $\pm$ 0.016 & 0.078 $\pm$ 0.01 & 0.085 $\pm$ 0.011 & 0.033 $\pm$ 0.023 & 0.066 $\pm$ 0.017 & 0.014 $\pm$ 0.006 & 0.028 $\pm$ 0.028 \\
HIP102152 & -0.126 $\pm$ 0.005 & -0.12 $\pm$ 0.007 & -0.11 $\pm$ 0.012 & -0.065 $\pm$ 0.009 & -0.078 $\pm$ 0.019 & -0.039 $\pm$ 0.013 & -0.016 $\pm$ 0.017 & -0.015 $\pm$ 0.009 & -0.001 $\pm$ 0.01 & 0.007 $\pm$ 0.019 & 0.012 $\pm$ 0.005 & -0.047 $\pm$ 0.027 \\
HIP10303 & 0.136 $\pm$ 0.005 & 0.123 $\pm$ 0.006 & 0.108 $\pm$ 0.014 & 0.084 $\pm$ 0.008 & 0.073 $\pm$ 0.006 & 0.076 $\pm$ 0.015 & 0.112 $\pm$ 0.012 & 0.09 $\pm$ 0.007 & 0.093 $\pm$ 0.016 & 0.115 $\pm$ 0.013 & 0.134 $\pm$ 0.005 & 0.099 $\pm$ 0.021 \\
... & ... & ... & ... & ... & ... & ... & ... & ... & ... & ... & ... & ... \\ 
\hline
\end{tabular}
\begin{tablenotes}
\item $\bf{Note.}$ The full version of this table is available online at the CDS.
\end{tablenotes}
\end{threeparttable}
\end{table}
\begin{table}
\centering
\caption{Stellar ages and masses determined for the sample of 79 solar twins.}
\begin{threeparttable}
\label{ages_Mv_logg}
\begin{tabular}{ccccccccccccc}
\hline
Star & Age$_{\rm mp}$ & Age$_{\rm ll-1\sigma}$ & Age$_{\rm ul-1\sigma}$ & Age$_{\rm ll-2\sigma}$ & Age$_{\rm ul-2\sigma}$ & Age$_{\rm mean}$ & Mass$_{\rm mp}$ & Mass$_{\rm ll-1\sigma}$ & Mass$_{\rm ul-1\sigma}$ & Mass$_{\rm ll-2\sigma}$ & Mass$_{\rm ul-2\sigma}$ & Mass$_{\rm mean}$ \\
 & (Gyr) & (Gyr) & (Gyr) & (Gyr) & (Gyr) & (Gyr) & (M$_{\odot}$) & (M$_{\odot}$) & (M$_{\odot}$) & (M$_{\odot}$) & (M$_{\odot}$) & (M$_{\odot}$) \\ \hline
HIP10175 & 3.1 & 2.8 & 3.5 & 2.4 & 4.4 & 3.2 $\pm$ 0.4 & 0.990 & 0.983 & 1.007 & 0.978 & 1.018 & 0.990 $\pm$ 0.002 \\
HIP101905 & 1.2 & 0.9 & 1.5 & 0.7 & 1.8 & 1.2 $\pm$ 0.3 & 1.080 & 1.073 & 1.097 & 1.067 & 1.108 & 1.080 $\pm$ 0.002 \\
HIP102040 & 2.4 & 2.0 & 2.8 & 1.6 & 3.1 & 2.4 $\pm$ 0.4 & 1.020 & 1.013 & 1.034 & 1.010 & 1.039 & 1.020 $\pm$ 0.002 \\
HIP102152 & 8.6 & 8.2 & 8.9 & 7.9 & 9.2 & 8.6 $\pm$ 0.3 & 0.980 & 0.972 & 0.997 & 0.962 & 1.022 & 0.978 $\pm$ 0.004 \\
HIP10303 & 5.9 & 5.5 & 6.3 & 5.0 & 6.6 & 5.8 $\pm$ 0.4 & 1.010 & 1.003 & 1.025 & 1.000 & 1.334 & 1.011 $\pm$ 0.003 \\
... & ... & ... & ... & ... & ... ... & ... & ... & ... & ... & ... & ... \\ 
\hline
\end{tabular}
\begin{tablenotes}
\item $\bf{Note.}$ The full version of this table is available online at the CDS.
\end{tablenotes}
\end{threeparttable}
\end{table}
\end{landscape}

\section{Discussion}
The chemical evolution of the Galactic components is often investigated through [X/Y]-[Fe/H] plots that give a picture of the progressive and differential enrichment of the X and Y elements and that can be used to calibrate the rate of star formation rates in the Galaxy and the yields from different progenitors (e.g., \citealt{Gilmore89,Reddy06,Shen15}). 
Behind the historical conventions, there are practical motivations that induced astronomers to study the [X/Y]-[Fe/H] distributions instead of investigating the direct dependence of the [X/Y] ratios with time. In fact, while stellar ages typically have large uncertainties, the iron abundance [Fe/H] is a quantity that can be more easily determined in stars and that can be used as a proxy of time. In addition, even if a [X/Y]-age relation gives immediate insights on the temporal evolution of the [X/Y] ratio within a stellar population \citep{Snaith15}, it is also true that the [X/Y]-[Fe/H] distributions contain a direct information on how the differential production of the X and Y elements varied with the metallicity of the progenitors or among populations with different chemical histories \citep{Kordopatis15,RojasArriagada16,Bekki17}. 
However, when studying the chemical evolution of the Galaxy, it is fundamental to bear in mind that [Fe/H] in a given population does not always increases with time \citep{Bensby07,Bensby14,Haywood13}. Also, the age-metallicity relation can be different at different Galactocentric distances. However, the effect of radial migration should not have a strong impact on our results. In fact, according to the data provided by the Geneva-Copenhagen Survey \citep{Nordstrom04,Holmberg09}, the stars in this sample have similar mean Galactocentric distances R$_{\rm m}$, defined as the mean value of the peri- and apo-galactocentric distances: the average value of R$_{\rm m}$ in the sample is 7.5~kpc with a standard deviation of 0.7~kpc. If we assume that the R$_{\rm m}$ is a good proxy of the Galactocentric distance where the stars were formed, then we can conclude that the stars in our sample were born within a restricted range of Galactocentric distances compared to the typical variation of the [X/Fe] ratios with R$_{\rm m}$ predicted by models \citep{Magrini09,Magrini17}.

%However, an increase in [Fe/H] in a given population does not always coincide with increasing time \citep{Bensby07,Haywood13}, therefore it is fundamental to bear in mind, when studying the chemical evolution of the Galaxy, that 
%the abundance ratios of elements in stars are function of time

%Therefore, since an increase in [Fe/H] in a given population does not always coincide with increasing time \citep{Bensby07,Haywood13}, it is fundamental to bear in mind, when studying the chemical evolution of the Galaxy, that the distribution of elements in the Galaxy is function of time, but it also depends on how different populations formed and evolved. 

%Therefore, since the stellar age and [Fe/H] do not always coincide \citep{Bensby07,Haywood13}, when studying the chemical evolution of the Galaxy it should be always taken into account that the distribution of abundances in a stellar population is function of time (as shown by the [X/Y]-age plots) and of the range of progenitor's metallicities (as shown by the [X/Y]-[Fe/H] plots).

%It is also important to notice that the stars of our sample have been selected among a restricted range of metallicities (i.e., [Fe/H]=0.0$\pm$0.1~dex), thus the ISM from which they formed has been processed by progenitors with similar ranges of metallicities. 

It is also important to note that our sample has been selected among a restricted range of metallicities (i.e., $-$0.1$\lesssim$[Fe/H]$\lesssim$0.1). Namely, the gas and environments from which all these stars formed have experienced different enrichment paths in the [Fe/H]-age diagram, that terminated with [Fe/H]$\sim$0.0 at the time of stellar formation. The enrichment path of the gas that formed the oldest objects in our sample must have been shorter than the enrichment path that led to the youngest stars. Hence, one may think that these ``short'' and ``long'' enrichment paths have been driven by progenitors of different natures (e.g., channels of nucleosynthesis, metallicities, masses, etc.). 

The variety of the possible progenitors that contributed to the chemical enrichment of a given population likely depends on its star formation history. For instance, the thin disc experienced only a slight enrichment along its 8~Gyr of stellar formation, compared to the sharp evolution of the thick disc (e.g., \citealt{Haywood13,Haywood16,Snaith15}). This would suggest that the thin disc solar twins, regardless their ages, have formed by gas enriched by progenitors with solar and sub-solar metallicities, while thick disc solar twins have probably formed by material mostly enriched by metal-poor progenitors. On the other hand, the youngest thin disc stars have formed by gas polluted by AGB progenitors with a broader range of masses, including also the low-mass stars, compared to the oldest stars that have been mainly enriched by AGB progenitors of higher mass.

%However, the fact that these stars share the same [Fe/H] also indicates that the range of metallicities covered by their progenitors was approximately the same, regardless to the length of the enrichment path. This must be especially true for the stars belonging to the thin disc, which experienced only a slight enrichment along its 8~Gyr of stellar formation, compared to the sharp evolution of the thick disc (e.g., \citealt{Haywood13,Haywood16,Snaith15}). On the other hand, it is clear that the youngest stars have formed by gas polluted by progenitors with a broader range of masses, including also the low-mass stars, compared to the oldest stars that have been mainly enriched by progenitors of higher mass.

Another important corollary of our selection criteria is that the [X/Y]-age stellar distributions discussed in the present paper are illustrative of the chemical evolution of stars with solar metallicities: other studies that attempt to analyse stars in different metallicity ranges would likely observe different (or slightly different) [X/Y]-age distributions (e.g., \citealt{Feltzing17}). 

In this section we discuss how the abundances of $n$-capture elements Sr, Y, Zr, Ba, La, Ce, Pr, Nd, Sm, Eu, Gd, and Dy evolved with time and how this information can be used to get new insights on the history of the Galactic disc. We also examine the reliability of stellar clocks based on abundance ratios (i.e., [Y/Mg] or [Y/Al]) and the age precision that can be achieved through them.

  \begin{figure*}
\centering
	\includegraphics[width=17.5cm]{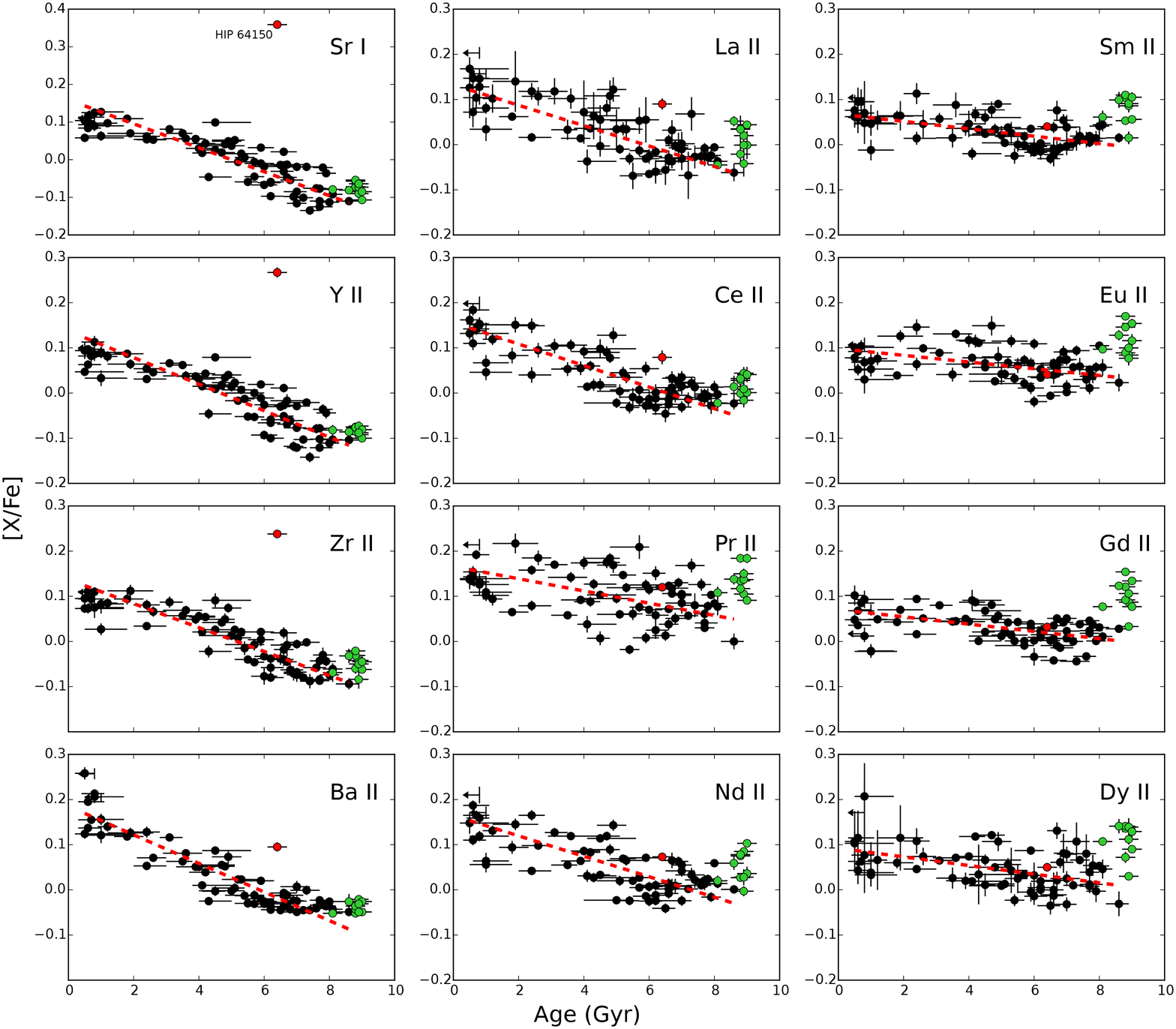}
    \caption{[X/Fe] ratios as a function of stellar ages. The black circles represent the thin disc stars, while the thick disc population is green. The red circle is the chemically anomalous star HIP64150. The [X/Fe]-age distributions have been fitted by the functions shown as red dashed lines, whose parameters are listed in Table~\ref{linearfit}.}
    \label{n_capture}
\end{figure*}

\subsection{Chemical evolution of the $n$-capture elements in the Galactic disc(s)}

The [X/Fe] vs age relations for the elements listed above are plotted in Fig.~\ref{n_capture}. We note that the [X/Fe] ratios of most of the species are highly correlated with the stellar ages and that each element is characterised by its own relation. 

Through the [X/Fe] ratios of the $\alpha$-elements, Bedell et al. (in prep) have identified 10 thick disc stars in our sample. This older population is plotted in Fig.~\ref{n_capture} with green circles, while the thin disc stars have been marked with black circles. We have also identified a star (i.e., HIP~64150; plotted in Fig.~\ref{n_capture} as a red circle) that is anomalously rich in the elements Sr, Y, Zr, and Ba in comparison to the bulk of thin disc stars. As detailed in Section~\ref{Spectroscopicanalysis}, the object is part of a binary system in which the main star is orbited by a white dwarf, thus its enhanced atmosphere may be the result of the pollution from an AGB companion (e.g., \citealt{Schirbel15}).

Interestingly, the ranges of [X/Fe] values covered by our sample of solar twins for each species are similar to the dispersions traced by solar analogs of [Fe/H]$\sim$0 observed by \citet{DelgadoMena17} in their [X/Fe]-[Fe/H] plots. This suggests that the dispersion of [X/Fe] values that they observe within the thin disc populations in the [X/Fe]-[Fe/H] plots results from the projection of the [X/Fe]-age relations sampled in different [Fe/H] bins. %In other words, the youngest stars should lie on the upper envelope of the [X/Fe]-[Fe/H] distribution of the thin disc, while the oldest should be those with the lowest [X/Fe] values.
In other words, the upper envelope of the [X/Fe]-[Fe/H] distribution traced by the thin disc should be populated by the youngest stars, while the oldest should be those in the lower envelope.

\begin{table*}
\centering
\caption{Results of the linear fitting of the [X/Fe]-age distributions traced by the thin disc population.}
\begin{threeparttable}[b]
\label{linearfit}
\begin{tabular}{cc|cccc|cc}
\hline
Element & $s$-process\tnote{*} & a & b & $\chi^{2}$ & $<$$\sigma$$_{[X/Fe]}$$>$ & p-value & D \\
 & [$\%$] & [10$^{-2}$ Gyr dex$^{-1}$] & [dex] & & & & \\ \hline
Sr & 68.9$\pm$5.9 & $-$3.2$\pm$0.2 & 0.159$\pm$0.011 & 3.94 & 0.03 & 6.4 10$^{-2}$ & 0.52 \\
Y & 71.9$\pm$8.0 & $-$2.94$\pm$0.18 & 0.137$\pm$0.010 & 3.15 & 0.03 & 1.5 10$^{-2}$ &0.62 \\
Zr & 66.3$\pm$7.4 & $-$2.65$\pm$0.16 & 0.137$\pm$0.009 & 2.38 & 0.03 & 1.7 10$^{-1}$ & 0.44 \\
Ba & 85.2$\pm$6.7 & $-$3.17$\pm$0.16 & 0.185$\pm$0.009 & 2.41 & 0.03 & 5.1 10$^{-1}$ & 0.32 \\
La & 75.5$\pm$5.3 & $-$2.3$\pm$0.2 & 0.132$\pm$0.012 & 2.49 & 0.03 & 1.3 10$^{-2}$ & 0.62 \\
Ce & 83.5$\pm$5.9 & $-$2.38$\pm$0.18 & 0.156$\pm$0.010 & 2.73 & 0.03 & 8.7 10$^{-2}$ & 0.49 \\
Pr & 49.9$\pm$4.3 & $-$1.4$\pm$0.3 & 0.166$\pm$0.015 & 8.15 & 0.04 & 1.5 10$^{-3}$ & 0.75 \\
Nd & 57.5$\pm$4.1 & $-$2.3$\pm$0.2 & 0.165$\pm$0.012 & 4.91 & 0.03 & 3.0 10$^{-2}$ & 0.57 \\
Sm & 31.4$\pm$2.2 & $-$0.82$\pm$0.18 & 0.068$\pm$0.010 & 2.17 & 0.02 & 5.6 10$^{-5}$ & 0.90 \\
Eu & 6.0$\pm$0.4 & $-$0.74$\pm$0.18 & 0.098$\pm$0.010 & 4.31 & 0.03 & 1.9 10$^{-4}$ & 0.85 \\
Gd & 15.4$\pm$1.1 & $-$0.80$\pm$0.16 & 0.071$\pm$0.009 & 4.61 & 0.02 & 5.6 10$^{-5}$ & 0.90 \\
Dy & 15.0$\pm$1.1 & $-$1.0$\pm$0.2 & 0.092$\pm$0.013 & 4.07 & 0.03 & 2.3 10$^{-3}$ & 0.72 \\ \hline
\end{tabular}
\begin{tablenotes}
\item [*] Solar $s$-process contribution percentages are from \citet{Bisterzo14}.
\end{tablenotes}
\end{threeparttable}
\end{table*}

  \begin{figure}
\centering
	\includegraphics[width=8.5cm]{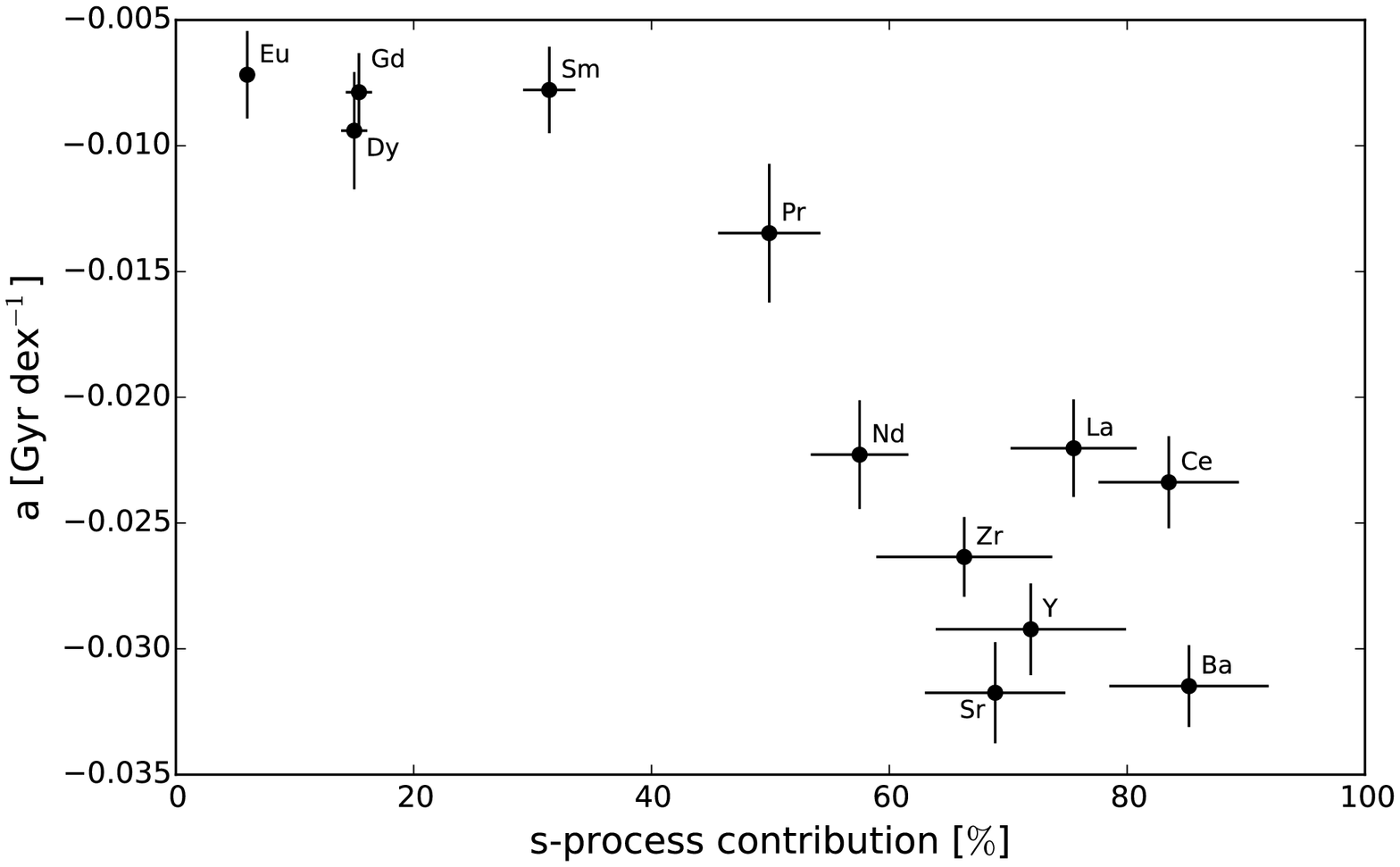}
    \caption{Slopes of the [X/Fe]-age distributions found for the $n$-capture elements (i.e., the ``a''  parameters in Table~\ref{linearfit}) as a function of their solar $s$-process contribution percentages taken from \citet{Bisterzo14}.}
    \label{scontr_slope}
\end{figure}

In general, thin disc stars are characterised by an increase of the [X/Fe] ratios as time goes on, and this is especially the case for $s$-process elements. The rise in $s$-process elements is likely the result of elemental synthesis in AGB stars. 
In fact, $n$-capture elements in the thin disc may have been produced mainly by low-mass AGB stars rather than by SNe, neutron star mergers or other nucleosynthetic sources  (see also \citealt{Battistini16}).

%In fact, during the evolution of the thin disc, the contribution to chemical evolution by AGB stars, mainly those of lower mass \citep{Busso01,Karakas16}, had an impact on the production $n$-capture elements that might has been greater than that of SNe, neutron star mergers or other sites of nucleosynthesis (see also \citealt{Battistini16}).} 

In order to verify this statement, we fitted the [X/Fe] - age distributions traced by the thin disc using the orthogonal distance regression method and the equation [X/Fe]=a$\times$Age+b. Due to its anomalous abundances, we have excluded from the fitting procedures the star HIP64150. The results of the linear fits are ove$r$-plotted in Fig. \ref{n_capture} and listed in Table~\ref{linearfit}, together with the solar $s$-process contribution percentages taken from \citet{Bisterzo14}, the $\chi$$^{2}$ of the fitting procedures and the average absolute deviation in ordinate $<$$\sigma$$_{[X/Fe]}$$>$. In Fig.~\ref{scontr_slope} we plot the [X/Fe]-age slopes of each element (i.e., the ``a'' parameters in Table~\ref{linearfit}) as a function of the contribution percentages from the $s$-process. The plot shows that a clear relation exists between the slopes and the $s$-process percentages: the elements with a lower contribution from the $s$-process, such as Eu, Gd and Dy, have flatter [X/Fe]-age distributions than the nearly-pure ``$s$-process'' elements like Ba, La, and Ce. In short, the production of $s$-process elements has been more active than that of $r$-process elements within the thin disc. A similar result has been previously found by \citet{Spina16b} on a smaller number of species and a smaller sample of solar twins. We interpret the relation in Fig. \ref{scontr_slope} as the confirmation that AGB stars are the main nucleosynthesis site responsible for the enrichment of $n$-capture elements during the thin disc phase.
%Therefore, the [X/Fe]-age distributions traced in Fig. \ref{n_capture} by the thin disc stars are mainly controlled by the rates of nucleosynthesis through the $s$-process, rather that those of the $r$-process.

In the last years, several high-resolution spectroscopic studies in open clusters have revealed that the [Ba/Fe] ratio is $\sim$0.6 for stars younger than 200 Myr decreasing to solar values for stars older than few Gyr \citep{DOrazi09c,DOrazi12,DeSilva13,Reddy15}. On the other hand, these studies also reported that high Ba abundances in young stars are not always accompanied by high abundances of the other $s$-process elements, such as Y, Zr, La, and Ce (see also \citealt{DOrazi17,Reddy17}). 
%On the other hand, these studies also reported that the other $s$-process elements exhibit nearly solar ratios, i.e. [X/Fe]$\sim$0.1-0.0, also at young ages (see also \citealt{DOrazi17}). 
The anomalous high [Ba/Fe] ratios in young cluster has been dubbed as \textit{the barium puzzle}, since models of Galactic chemical evolution are not able to reproduce the steep increase of [Ba/Fe] during the last Gyr \citep{Maiorca12,Mishenina15}. Recently, \citet{Reddy17} have noted that there is an increasing scatter in their [Ba/Fe]-age distribution with decreasing stellar ages, so that the stars younger than 1 Gyr have [Ba/Fe] values between 0.1 and 0.7 dex, in contrast to the clear [X/Fe]-age trends of the other heavy elements. 
%In addition, they also noted that high Ba abundances are not always accompanied by high abundances of the other $s$-process elements, such as La and Sm. 
Therefore, they have proposed that the enhancement of Ba might be unrelated to the Galactic chemical evolution, but it could be the result of an overestimation of Ba in relatively active photospheres, i.e. young stars. Namely, the Ba~II lines observed in the visible spectral range (e.g., 4554, 5853, 6141, and 6496 $\AA$) are formed in superficial layers that are more turbulent than the microturbulence value derived from LTE analysis of Fe lines formed in deeper layers. Thus, the assumption of a microturbulence derived from Fe would not be sufficient to reproduce the broadening on the Ba~II lines, which would result in a severe overestimation of the Ba abundance. At odds with the observational results discussed by \citet{Reddy17}, the [Ba/Fe]-age distribution that we got from our data is tight in comparison with the distributions obtained for the other elements (see the $\chi^{2}$ values in Table~\ref{linearfit}) and its scatter around the linear fit does not significantly increase at smaller ages. In addition, from our analysis all the solar twins have [Ba/Fe]$<$0.3. Most interestingly, the [Ba/Fe]-age slope is in excellent agreement with the tendency outlined by the other elements in Fig.~\ref{scontr_slope}. This indicates that higher Ba yields go with higher yields of the other heavy elements produced through the $s$-process. In other words, our analysis is not affected by the systematics discussed by \citet{Reddy17}. The reliability of our results can probably be ascribed to the fact that we have analysed twin stars in a differential fashion. This has likely reduced (or cancelled out) the effect of any systematic error on the determinations of stellar parameters and abundances. In fact, it should be noted that \citet{Nissen16}, through a differential analysis of solar twin stars, obtained a [Ba/Fe]-age distribution that closely resembles that plotted in Fig.~\ref{n_capture}. A similar distribution was also found by \citet{Reddy17} for their sample of solar twins (see their Fig. 5), with the exception of two outliers HD~42807 and HD~59967 (i.e., HIP~29525 and HIP~36515). Interestingly, these two stars are also part of our sample, but their [Be/Fa] ratios (i.e., 0.206 and 0.258) and their ages (i.e., 0.8 and 0.5 Gyr) determined in this work nicely agree with our linear fit. Also, our selection of a linelist that includes both strong and weak Fe lines must have contributed in a solid determination of the stellar parameters.

In Fig.~\ref{n_capture} it is also seen that the thick disc population does not necessarily follow the main distributions traced by the thin disc. For example, there is a discontinuity between the thin and the thick disc in the $r$-process elements Eu, Gd and Dy: the thick disc appears strongly enhanced in these elements relative to the oldest stars in the thin disc. Interestingly, this behaviour is similar to that found for the $\alpha$-elements \citep{Haywood13,Nissen15,Spina16,Spina16b}, which are tracers of massive star formation bursts. 
Recently, \citet{Snaith15} and \citet{Haywood16}, by studying the evolution of the $\alpha$-elements in the thin and thick discs, concluded that the star formation rate in the thick disc was three times more intense than that in the thin disc. In fact, while the thick disc has undergone a burst of star formation that consumed most of the gas available in the disc, the thin disc phase is compatible with a constant star formation rate. These authors also measured a dip in the star formation rate that lasted $\sim$1Gyr during the transition between the thick and thin discs. Whatever the process re-initiated that formation activity in the disc (e.g., external gas accretion, the cooling of the disc after the first star formation burst), during the quiescent phase of star formation, the disc's gas has been strongly pre-enriched in Fe by Type Ia SNe of the thick disc population, but less in $\alpha$-elements. This generated the clear discontinuity between the thin and thick discs visible in the [$\alpha$/Fe]-age distribution and in the [$\alpha$/Fe]-[Fe/H] diagrams \citep{Masseron15}. The fact that there is a discontinuity between the thin and thick disc populations traced by $r$-process elements in both the [X/Fe]-age and [X/Fe]-[Fe/H] plots \citep{Reddy06}, suggests that a variety of sites, including core-collapse SNe, may produce $r$-process elements. Massive compact star mergers, which are thought to be the main producers of $r$-process elements, have delay times longer than those of Type II SNe and comparable to those of Type Ia SNe or longer \citep{Maoz14,Mennekens16,Cote17}. Therefore, considering these mergers as unique channels of $r$-process element production would not explain the similarity between r-process and $\alpha$-elements. On the other hand, it would indicate that these species have shared the same sites of nucleosynthesis, at least until the early thin disc phase.

We also note that the thick disc population approximately shares the same [Ba/Fe] values of the oldest stars of the thin disc, indicating that Ba and Fe have been produced in similar quantities during the quiescent phase. In Table~\ref{linearfit}, we list the Kolmogorov-Smirnov (K-S) statistic D and its associated p-value from a comparison of the oldest (i.e., age$>$7 Gyr) thin disc and thick disc populations for each element. In general, the smaller the contribution from the $s$-process is, the more prominent the discontinuity between the two populations appears to be. This occurs because the $s$-process elements, differently from the $\alpha$ species, are not tracers of intensive bursts of star formation.

\subsection{Light and heavy $s$-process elements}
\label{h$s$-ls}
\begin{figure*}
\centering
	\includegraphics[width=17.5cm]{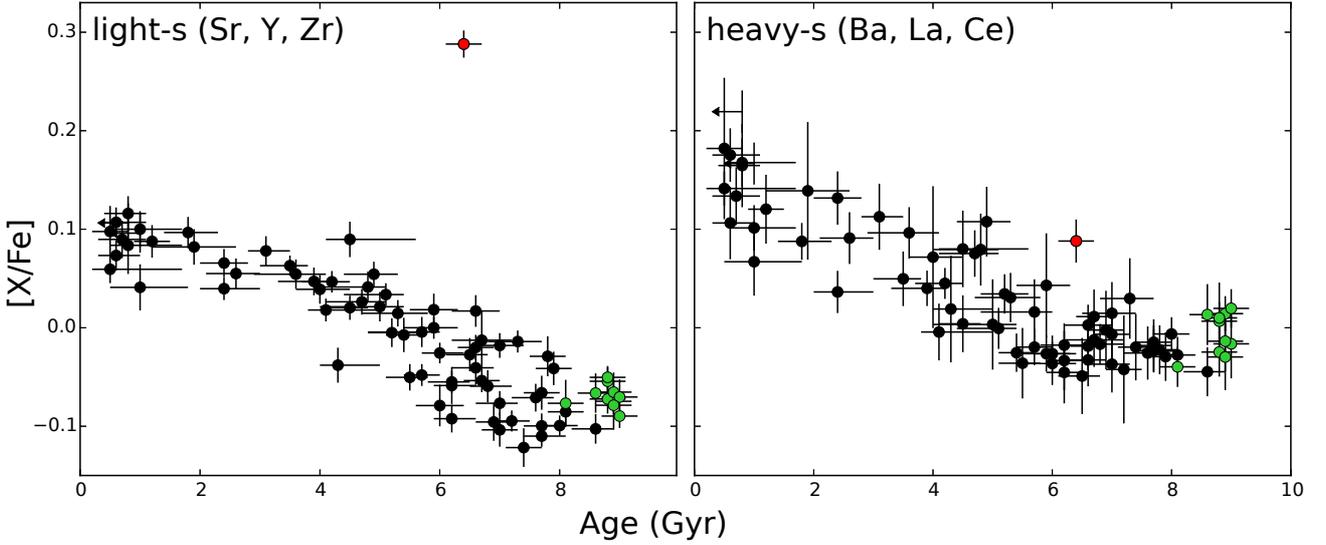}
    \caption{The panels show the [ls/Fe]-age (left) and [hs/Fe]-age (right) relations, where $ls$ and $hs$ are the averages of the abundances of the light elements (Sr, Y, and Zr) and of the heavy elements (Ba, La, Ce), respectively. The symbols are the same of Fig.~\ref{n_capture}.}
    \label{ls_hs_Fe}
\end{figure*}

In Fig. \ref{ls_hs_Fe} we show [X/Fe]-age plots for the light ($ls$) and heavy $s$-process elements ($hs$), where [ls/Fe]=([Sr/Fe]+[Y/Fe]+[Zr/Fe])/3 and [hs/Fe]=([Ba/Fe]+[La/Fe]+[Ce/Fe])/3. The distributions of the two classes of elements present a slightly different trait: while the [hs/Fe] ratio experiences a constant increment as the time goes on, the [ls/Fe]-age distribution flattens at ages $\lesssim$4 Gyr. This evidence is clearly visible also in Fig.~\ref{n_capture} where the majority of the stars younger than 3~Gyr have Sr, Y, and Zr abundances that are slightly lower than those predicted by the linear fit. Since the bulk of the $ls$ and $hs$ elements is produced through the $s$-process with percentages that in averages do not differ more than a 10$\%$ (see Table~\ref{linearfit}), this effect is likely related to the $s$-process itself and not due to other channels of nucleosynthesis, such as the $r$-process.

Theoretical calculations of AGB nucleosynthesis predict that the $^{13}$C pocket is the dominant source of free neutrons for low-mass ($\leq$3 M$_{\odot}$) AGB stars \citep{Gallino98,Lugaro03,Cristallo09}. The high flux of neutrons generated by this source permits a greater production of $hs$ elements at the expenses of the $ls$ elements \citep{Karakas16}. Since low-mass stars contribute at later stages to the Galactic chemical evolution, the flattening of the [ls/Fe]-age distribution at 4~Gyr observed in Fig. \ref{ls_hs_Fe} could be related to the positive [hs/ls] ratios yielded by the $s$-process in the $^{13}$C pocket. In addition, since, the neutron flux from the $^{13}$C pocket is highly dependent on the stellar metallicity \citep{Busso01}, one would expect to observe a large dispersion in the youngest branch of the [ls/Fe]-age distribution. On the contrary, the [ls/Fe]-age distribution of youngest stars ($\lesssim$4 Gyr) seems tighter than that of the oldest. This probably indicates that the gas from which the youngest solar twins have formed has been enriched by low-mass AGB stars of similar metallicities or that the mixing of the ISM has been more efficient at later times.

%As we mentioned in Section~\ref{intro}, theoretical calculations of AGB nucleosynthesis predict that the [hs/ls] ratio of the AGB products is $\sim$0 for AGB stars of solar metallicities and masses $\geq$3 M$_{\odot}$, but it increases as the mass of the AGB star diminish \citep{Karakas16}. This occurs because a region called $^{13}$C pocket, that is a primary source of free neutrons for low-mass stars, develops in stars of $\sim$3M$_{\odot}$ and is assumed to be larger for AGB stars with lower masses \citep{Gallino98,Lugaro03,Cristallo09}. Therefore, a higher flux of free neutrons permits a greater production of hs elements at the expenses of the ls elements. Since stars of lower masses contribute at later stages to the Galactic chemical evolution, the flattening of the [ls/Fe]-age distribution at 4~Gyr observed in Fig. \ref{ls_hs_Fe} could be related to the different regime of nucleosynthesis due to the emergence of the $^{13}$C pocket in low-mass AGB stars. In addition, since, the neutron flux from the $^{13}$C pocket is highly dependent from the stellar metallicity \citep{Busso01}, one would expect to observe a large dispersion in the youngest branch of the [ls/Fe]-age distribution. On the contrary, the [ls/Fe]-age distribution of youngest stars ($\lesssim$4 Gyr) seems tighter than that of the oldest. This probably indicates that the gas, from which the youngest solar twins of our sample have formed, has been enriched by low-mass AGB stars of similar metallicities. 

\begin{figure}
\centering
	\includegraphics[width=8.5cm]{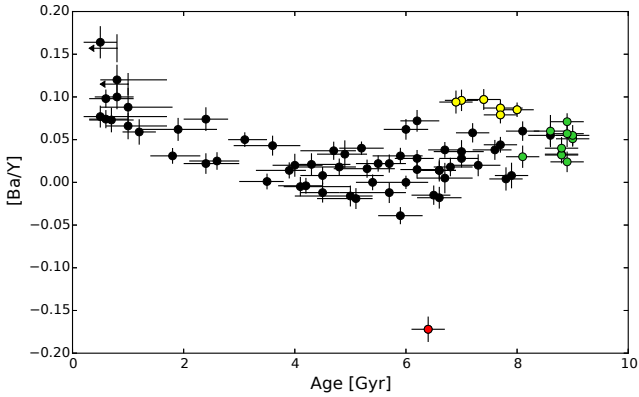}
    \caption{[Ba/Y] as a function of the stellar ages. The yellow dots are the old thin disc stars with high [Ba/Y] ratios. The other symbols are same of Fig.~\ref{n_capture}.}
    \label{Ba_Y}
\end{figure}

The ``$s$-process relative indicators'' are abundance ratios of two or more elements mainly produced in AGB stars. Differently from the [s/Fe] ratios, where Fe is not produced 
through the $s$-process, these relative indicators, in a first approximation, do not depend from variables, such as the mass loss, the stellar lifetimes, and mixing process, that are highly uncertain in AGB models \citep{Lugaro12}. For this reason, they provide independent insight to the study of the elemental nucleosynthesis of AGB stars. In Fig~\ref{Ba_Y} we show the [Ba/Y] ratio as a function of the stellar age. Since Y and Ba belong to the light and heavy $s$-process elements, respectively, the plot in Fig~\ref{Ba_Y} illustrates how the relative abundances of $hs$ and $ls$ vary with time in solar twin stars. We observe that the stars younger than 6~Gyr experience a [Ba/Y] increase as the time goes, which is in agreement with the positive [hs/ls] yielded by low-mass stars of solar and sub-solar metallicities. %We also noted that the scatter around the mean for stars younger than 6~Gyr is $\sim$0.05, while the thin disc stars with ages of 6-8 Gyr are more dispersed. 
However, there is an inversion of this trend at $\sim$6 Gyr, with an increment of the [Ba/Y] dispersion for the oldest thin disc stars. This behaviour might be connected to the dependence of the AGB stars yields on metallicity. During the $s$-process, the number of atoms that can capture free neutrons is proportional to the stellar metallicity Z, thus the flux of free neutrons is expected to grow when Z decreases \citep{Clayton88}. As a consequence, the production of heavier elements is more favourable at lower metallicities \citep{Busso01,Lugaro12,Fishlock14,Cristallo15}. Namely, AGB models predict that the [hs/ls] ratio in the winds of a 3M$_{\odot}$ AGB star is $-$0.026 at Z=Z$_{\odot}$ and +0.320 at Z=1/2Z$_{\odot}$ \citep{Karakas16}. Hence, it is plausible that the old (6-8 Gyr) stars with highest [Ba/Y] ratios (yellow points in Fig~\ref{Ba_Y}) were likely contaminated by lower metallicity AGB winds.

%The inversion of the trend observed at $\sim$6 Gyr in the [Ba/Y]-age distribution and the higher dispersion of the oldest stars may be connected to the dependence of the AGB stars yields on metallicity. In fact, since the neutrons are captured by heavy-elements, whose number is proportional to the stellar metallicity Z, then the neutron flux increases when Z decreases \citep{Clayton88}. This implies that the production of heavier elements is more favourable at lower metallicities \citep{Busso01,Lugaro12}. Namely, AGB models predict that the [hs/ls] ratio in the winds of a 3M$_{\odot}$ AGB star is $-$0.026 at Z=Z$_{\odot}$ and +0.320 at Z=1/2Z$_{\odot}$ \citep{Karakas16}. Hence, it is plausible that the old (6-8 Gyr) stars with highest [Ba/Y] ratios (yellow points in Fig~\ref{Ba_Y}) are those that have been formed by gas mostly polluted by metal-poor AGB stars from the early thin disc or from previous Galactic populations, such as the thick disc. 

  \begin{figure}
\centering
	\includegraphics[width=8.5cm]{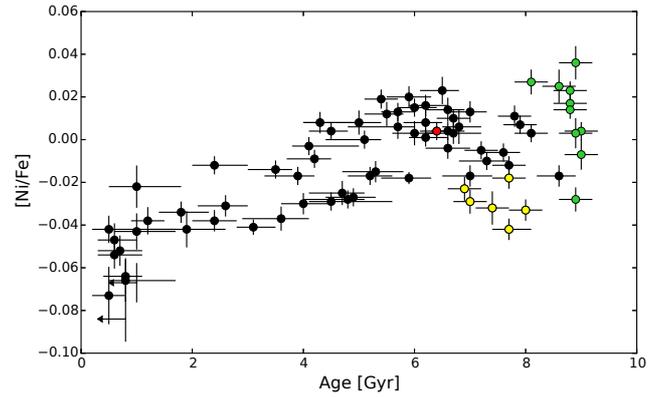}
    \caption{[Ni/Fe] as a function of the stellar ages. The Ni abundances are from Bedell et al. (in prep). The yellow dots are the old thin disc stars with high [Ba/Y] ratios (see Fig.~\ref{Ba_Y}). The other symbols are same of Fig.~\ref{n_capture}.}
    \label{Ni_Fe}
\end{figure}

In this regard, it is also remarkable to notice that the high-[Ba/Y] stars in Fig~\ref{Ba_Y} are also those with low [Ni/Fe] ratios in Fig~\ref{Ni_Fe}. As suggested by \citet{Nissen16}, the old population (age $\gtrsim$6 Gyr) with low [Ni/Fe] ratios were formed in regions with lower metallicity than the regions where the old stars with high [Ni/Fe] were formed. Therefore, the scenario proposed by Nissen could also explain the existence of an old group of thin disc stars with high [Ba/Y] ratios. This scenario would also imply that the ISM was not efficiently mixed shortly after the quiescent phase and during the early stage of the thin disc, but that this mixing likely has become more effective with time.

\begin{figure*}
\begin{center}
\includegraphics[width=17.5cm]{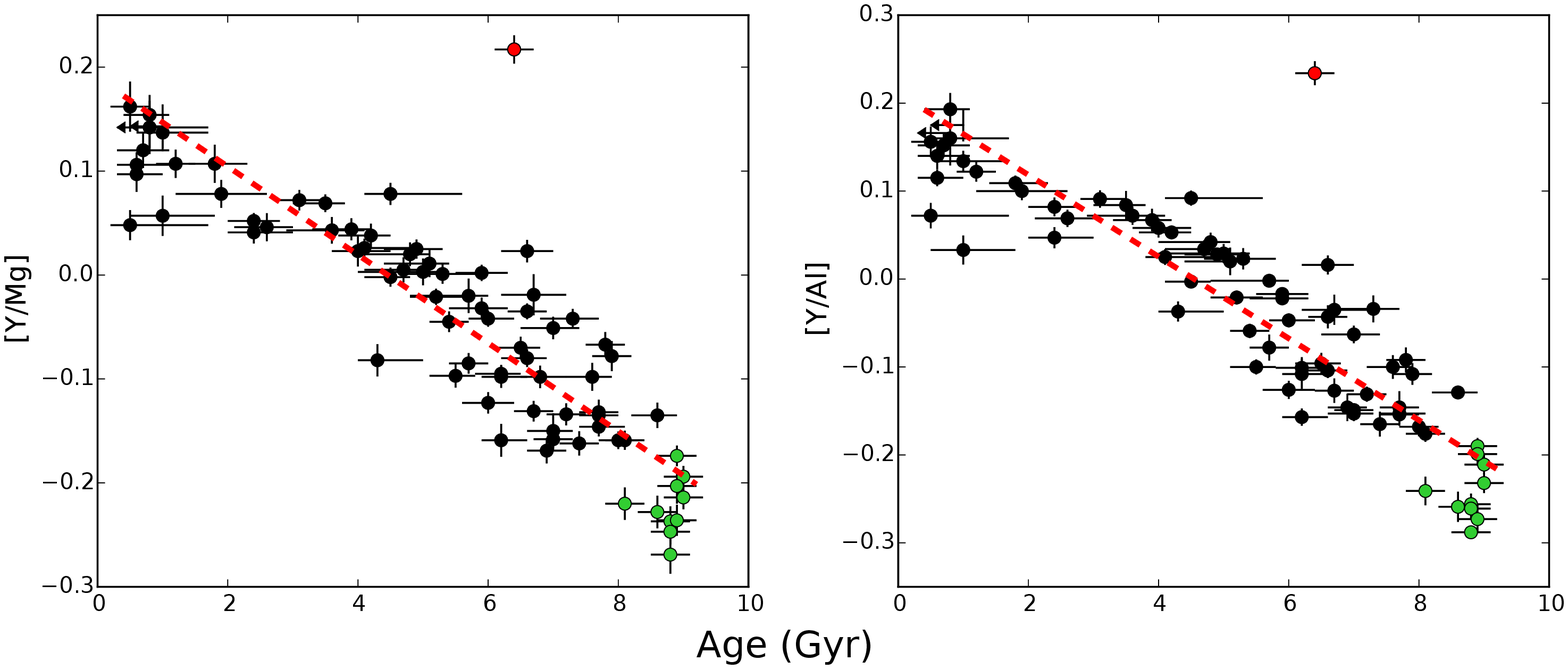}
\caption{[Y/Mg] and [Y/Al] as a function of age, with the same symbols of Fig. \ref{n_capture}. The red dashed lines correspond to the linear functions described by Eq. \ref{Y_Mg} and \ref{Y_Al}, while the blue solid lines represent the quadratic functions in Eq. \ref{Y_Mg_quad} and \ref{Y_Al_quad}.}
\label{chemical_clocks}
\end{center}
\end{figure*}

\subsection{Chemical clocks}
The [Y/Mg] and [Y/Al] ratios are sensitive indicators of stellar ages for solar twin stars, due to their tight and steep dependence on time \citep{DaSilva12,Nissen15,TucciMaia16,Spina16b}. These indicators are a natural consequence of Galactic chemical evolution, which proceeded through different channels of nucleosynthesis with different timescales. Namely, while Mg and Al are mainly produced by core-collapse SNe, which enriched the medium in short timescales \citep{Matteucci14}, the bulk of Y in the Galaxy has been synthesised by 1-8 M$_{\odot}$ stars, once they have reached the AGB phase (\citealt{Nomoto13,Karakas14} and references therein). 

The use of the [Y/Mg] ratio as a chemical clock has been initially proposed by \citet{Nissen15,Nissen16} and then subsequently confirmed by \citet{TucciMaia16} on an independent sample of solar twin stars. More recently, \citet{Spina16b} showed that the [Y/Al] ratio has a steeper dependence on stellar age than [Y/Mg], thus it would represent an even more precise chemical clock. Furthermore, \citet{Slumstrup17} showed that the [Y/Mg] clock can also be used for solar-metallicity giants. 

As pointed out by \citet{Feltzing17}, these chemical clocks can only be employed to infer the ages of solar-metallicity stars: since the production of Y decreases at lower metallicities, the correlations between [Y/Mg] or [Y/Al] and age would flatten for stars with metallicities lower than solar.

Fig. \ref{chemical_clocks} shows the [Y/Mg] and [Y/Al] ratios as a function of stellar age. The linear fits that include both the thin and thick disc stars, with the exception of the anomalous Y-rich star (red dot), and which take into account the errors in both the coordinates, result in the following relations:

\begin{equation}
\label{Y_Mg}
\left[ \frac{\rm Y}{\rm Mg} \right]=0.204\left(\pm0.014\right)-0.046\left(\pm0.002\right)\times \rm Age\left[\rm Gyr \right],\\
\end{equation}
\begin{equation}
\label{Y_Al}
\left[ \frac{\rm Y}{\rm Al} \right]=0.231\left(\pm0.014\right)-0.051\left(\pm0.002\right)\times \rm Age\left[\rm Gyr \right].
\end{equation}

In Table~\ref{clock} we compare the slopes of the two relations above determined by different authors through high-precision spectroscopic analysis of solar twin stars. We can note that the slopes found by our analysis are significantly more negative than those obtained in previous studies. The difference between our determinations and those of \citet{TucciMaia16} and \citet{Spina16b} may be related to the correction that we applied to the grid of isochrones in order to take into account the $\alpha$-enhancement of the model atmospheres (see Section~\ref{ages_masses}). This correction, neglected in our previous works, reduced the ages of the oldest stars by $\sim$1 Gyr (see Fig~\ref{comparison_alpha}). On the other hand, even if Nissen considered the impact of the $\alpha$-abundances in the age determinations, his sample includes only three thick disc stars which are on average 0.6 Gyr older than the mean age of the thick disc stars of our sample. This could account for the different slopes. In fact, if we perform a linear fitting of the thin disc distribution alone (see Table~\ref{clock}, last column), the resulting slopes are consistent within the errors with those of \citet{Nissen15}, \citet{TucciMaia16} and \citet{Spina16b}. 

\begin{table*}
\centering
\caption{Slopes of the [Y/Mg]-age and [Y/Al]-age distributions obtained through high-precision spectroscopic analysis on samples of solar twin stars.}
\label{clock}
\begin{tabular}{ccccccc}
\hline
Slope & \citet{Nissen15} & \citet{TucciMaia16} & \citet{Spina16b} & \citet{Nissen16} & Present work & Present work \\
(dex Gyr$^{-1}$) & (no thick disc) & & & & & (no thick disc) \\ \hline
N. stars & 18 & 88 & 41 & 21 & 76 & 66 \\ 
$\rm [$Y/Mg] & $-$0.0404$\pm$0.0019 & $-$0.041$\pm$0.001 & $-$0.0404$\pm$0.0019 & $-$0.0371$\pm$0.0013 & $-$0.045$\pm$0.002 & $-$0.042$\pm$0.003 \\
$\rm [$Y/Al] & --- & --- & $-$0.0459$\pm$0.0018 & $-$0.0427$\pm$0.0014 & $-$0.051$\pm$0.002 & $-$0.046$\pm$0.002 \\ \hline
\end{tabular}
\end{table*}

The average scatter of the two chemical clocks are 1.0 Gyr for [Y/Mg] and 0.9 Gyr for [Y/Al]. Considering that the typical uncertainties of our ages determinations is 0.4~Gyr and including the thick disc stars, the intrinsic scatter in age around the relations described by Eq. \ref{Y_Mg} and \ref{Y_Al} is 0.9~Gyr. However, as we observed in Section~\ref{h$s$-ls}, the increment of the Y abundance with time flattens for ages $\lesssim$3 Gyr. This behaviour clearly affects also the temporal evolution of the [Y/Mg] or [Y/Al] ratios that are not perfectly linear too. As a consequence, the linear fittings tend to overestimate the ages of stars younger than 3~Gyr by 0.8-0.9~Gyr. 

By fitting the distributions of thin and thick discs with quadratic functions, we found the following solutions:
%\begin{equation}
%\label{Y_Mg_quad}
%\small
%\left[ \frac{\rm Y}{\rm Mg} \right] = 0.125\left(\pm0.017\right) - 0.007\left(\pm0.009\right)\times \rm Age- 0.0038\left(\pm0.0011\right)\times \rm Age^2,\\
%\end{equation}
%\begin{equation}
%\label{Y_Al_quad}
%\left[ \frac{\rm Y}{\rm Al} \right]=0.150\left(\pm0.015\right)-0.0115\left(\pm0.009\right)\times \rm Age- 0.0038\left(0.0011\right)\times \rm Age^2.
%\end{equation}

\begin{equation}
\label{Y_Mg_quad}
\begin{split}
\left[ \frac{\rm Y}{\rm Mg} \right] = & 0.125\left(\pm0.017\right) - 0.007\left(\pm0.009\right)\times \rm Age \\
 & -0.0038\left(\pm0.0011\right)\times \rm Age^2,\\
\end{split}
\end{equation}

\begin{equation}
\label{Y_Al_quad}
\begin{split}
\left[ \frac{\rm Y}{\rm Al} \right]= & 0.150\left(\pm0.015\right)-0.0115\left(\pm0.009\right)\times \rm Age\\
 & -0.0038\left(0.0011\right)\times \rm Age^2.
\end{split}
\end{equation}

From these higher-order functions, we find that the average scatter is reduced to 0.5~Gyr for both [Y/Mg] and [Y/Al], which is very close to the typical uncertainties of our age determinations (i.e., 0.4~Gyr).

\section{Conclusions}
We used high-quality HARPS spectra of 79 solar twins to obtain precise estimates of their atmospheric parameters (T$_{\rm eff}$, log~g, [Fe/H], and $\xi$) and abundances of 12 neutron-capture elements (Sr, Y, Zr, Ba, La, Ce, Pr, Nd, Sm, Eu, Gd, and Dy). The knowledge of the stellar parameters and absolute magnitudes of these stars allowed us to determine their stellar ages through the isochrone method with typical precisions of 0.4~Gyr.

Our main scientific results can be summarised as follows:

\begin{itemize}
\item The thin and thick disc populations, identified through the $\alpha$-elements abundances, are separated in the [X/Fe] vs age plots showed in Fig.~\ref{n_capture}. The [X/Fe] ratios of the thin disc distribution are highly correlated with stellar age for the $s$-process elements (e.g., Ba), while the [X/Fe]-age distributions of the $r$-process elements (e.g., Eu) are nearly flat. Indeed, the slopes of the [X/Fe]-age distributions are dependent on the $s$-process contribution percentage of the X element (see Fig.~\ref{scontr_slope}). This indicates that, the enrichment from AGB stars is the predominant channel of nucleosynthesis of $n$-capture elements during the evolution of the thin disc. Despite the flat slopes traced by the thin disc stars, the nearly pure $r$-process elements Eu, Gd, and Dy appear more enhanced in the thick disc population. This dichotomy between the thin and thick discs resembles the [X/Fe]-age distributions outlined by the $\alpha$-elements, which are more abundant in the thick disc than in the thin disc due to an intensive nucleosynthesis by Type~II SNe during the early stages of the Galactic disc. The sites of production of the $r$-process elements are not very well known at present; however it is likely that they are mainly synthesised during the mergers of massive compact objects. On the other hand, the similarity to the distributions of the $\alpha$-elements suggests that the $r$-process elements are also produced by Type~II SNe and that they could be additional tracers of intensive star formation episodes. There is a possible direct observational test that could be performed on OB associations which are regarded as a proto-type of triggered star formation, such as Orion or Sco-Cen. If Type~II SNe are producing copious $r$-process elements, then the youngest sub-clusters should be enriched in these elements relative to the oldest stars \citep{Cunha92}.

\item The [X/Fe]-age distributions of the thin disc stars have been fitted by linear functions. The parameters resulting from this linear fitting, listed in Table~\ref{linearfit}, outline the progressive enrichment of the Galactic disc in $n$-capture elements and can also be employed to identify chemical anomalous stars heavily enhanced by evolved companions, such as HIP64150.

\item The [Ba/Fe] ratios determined in young open clusters and stellar associations are $\sim$0.6. These extreme values are likely the consequence of systematics in the spectral analysis that lead to a serious overestimation of the Ba abundances in these young stars \citep{Reddy17}. However, the temporal evolution of the [Ba/Fe] ratio determined from our analysis is in agreement with the [X/Fe]-age distributions of the other elements (see Fig.~\ref{scontr_slope}): the higher Ba abundances of the youngest stars (typically [Ba/Fe]$\sim$0.2-0.1) are accompanied by higher abundances of the other heavy elements produced through the $s$-process. Therefore, our results indicate that the differential analysis of twin stars is not significantly affected by any systematic overestimations of the Ba abundances.

\item We have shown that the [ls/Fe]-age distribution flattens at ages $\lesssim$4~Gyr in contrast to the [hs/Fe] ratios that show a steady increase as time goes on (see Fig.~\ref{ls_hs_Fe}). This difference in the two distributions could be related to the nucleosynthesis of low-mass AGB stars, which preferentially release into the ISM heavy $s$-process elements at the expense of the light $s$-process elements. This characteristic of the low-mass AGB stars also explains the [Ba/Y] decrement with time observed in Fig.~\ref{Ba_Y} for the stars younger than 6~Gyr. However, this trend is not followed by the oldest thin disc stars, which, on average, have higher [Ba/Y] ratios. It is likely that the 6-8~Gyr old stars with highest [Ba/Y] ratios were formed in more metal poor environments.% If this hypothesis is confirmed by further studies, the high [Ba/Y] ratios of these stars could be an additional hint of a continuity between the thin and thick discs.

\item We have calibrated the [Y/Mg] and [Y/Al] chemical clocks over a sample of 76 solar twin stars belonging to the thin and thick discs populations. Using the linear relations expressed by Eq. \ref{Y_Mg} and \ref{Y_Al}, it is possible to estimate the ages of solar twin stars with uncertainties of $\sim$0.8~Gyr, while the quadratic functions described by Eq. \ref{Y_Mg_quad} and \ref{Y_Al_quad} allow to reach a precision that is similar to the typical precision of our age determinations.

\end{itemize}

We have shown that precise abundances of solar twin stars can provide fundamental insights on the formation and evolution of the Galactic discs. In addition, since the chemical enrichment of gas in our Galaxy can follow different paths in the [X/Fe]-[Fe/H]-age space, through the analysis of the [X/Y]-ages relations we can achieve a more comprehensive understanding of the distributions of stellar populations in the [X/Y]-[Fe/H] plots, and \textit{vice versa}. The works of \citet{Haywood13,Haywood15,Haywood16,Snaith15} also express this possibility, even if they lack the high precision that is achievable through the differential analysis of twin stars. Our work and other studies \citep{Nissen15,Nissen16,Spina16b} exploit this high precision in the determination of abundances ($\sigma$$<$0.01~dex) and ages ($\sigma$$<$1~Gyr), but only of stars with solar metallicity (i.e., $-$0.10$\leq$[Fe/H]$\leq$+0.10). A significant progression in the knowledge of the chemical evolution of the Galaxy will be possible by the analysis of the [X/Y]-age relations for stars with metallicities different than solar. This would allow us to slice the [X/Y]-[Fe/H] plane at constant [Fe/H] and study, for each [Fe/H] bin, the temporal evolution of the [X/Y] ratios. A further fundamental step will be provided by theoretical models of chemical evolution that are now challenged to reproduce the [X/Y]-age relations.

\section*{Acknowledgements}

LS and JM acknowledge the support from FAPESP (2014/15706-9 and 2012/24392-2). MA has been supported by the Australian Research Council (grant FL110100012).
%%%%%%%%%%%%%%%%%%%%%%%%%%%%%%%%%%%%%%%%%%%%%%%%%%



\bibliographystyle{mnras}
\bibliography{/Users/lspina/Dropbox/papers/bibliography.bib} % if your bibtex file is called example.bib

\begin{thebibliography}{}
\makeatletter
\relax
\def\mn@urlcharsother{\let\do\@makeother \do\$\do\&\do\#\do\^\do\_\do\%\do\~}
\def\mn@doi{\begingroup\mn@urlcharsother \@ifnextchar [ {\mn@doi@}
  {\mn@doi@[]}}
\def\mn@doi@[#1]#2{\def\@tempa{#1}\ifx\@tempa\@empty \href
  {http://dx.doi.org/#2} {doi:#2}\else \href {http://dx.doi.org/#2} {#1}\fi
  \endgroup}
\def\mn@eprint#1#2{\mn@eprint@#1:#2::\@nil}
\def\mn@eprint@arXiv#1{\href {http://arxiv.org/abs/#1} {{\tt arXiv:#1}}}
\def\mn@eprint@dblp#1{\href {http://dblp.uni-trier.de/rec/bibtex/#1.xml}
  {dblp:#1}}
\def\mn@eprint@#1:#2:#3:#4\@nil{\def\@tempa {#1}\def\@tempb {#2}\def\@tempc
  {#3}\ifx \@tempc \@empty \let \@tempc \@tempb \let \@tempb \@tempa \fi \ifx
  \@tempb \@empty \def\@tempb {arXiv}\fi \@ifundefined
  {mn@eprint@\@tempb}{\@tempb:\@tempc}{\expandafter \expandafter \csname
  mn@eprint@\@tempb\endcsname \expandafter{\@tempc}}}

\bibitem[\protect\citeauthoryear{{Amelin}, {Kaltenbach}, {Iizuka}, {Stirling},
  {Ireland}, {Petaev}  \& {Jacobsen}}{{Amelin} et~al.}{2010}]{Amelin10}
{Amelin} Y.,  {Kaltenbach} A.,  {Iizuka} T.,  {Stirling} C.~H.,  {Ireland}
  T.~R.,  {Petaev} M.,   {Jacobsen} S.~B.,  2010, \mn@doi [Earth and Planetary
  Science Letters] {10.1016/j.epsl.2010.10.015}, \href
  {http://adsabs.harvard.edu/abs/2010E%26PSL.300..343A} {300, 343}

\bibitem[\protect\citeauthoryear{{Argast}, {Samland}, {Thielemann}  \&
  {Qian}}{{Argast} et~al.}{2004}]{Argast04}
{Argast} D.,  {Samland} M.,  {Thielemann} F.-K.,   {Qian} Y.-Z.,  2004, \mn@doi
  [\aap] {10.1051/0004-6361:20034265}, \href
  {http://adsabs.harvard.edu/abs/2004A%26A...416..997A} {416, 997}

\bibitem[\protect\citeauthoryear{{Battistini} \& {Bensby}}{{Battistini} \&
  {Bensby}}{2016}]{Battistini16}
{Battistini} C.,  {Bensby} T.,  2016, \mn@doi [\aap]
  {10.1051/0004-6361/201527385}, \href
  {http://adsabs.harvard.edu/abs/2016A%26A...586A..49B} {586, A49}

\bibitem[\protect\citeauthoryear{{Bedell}, {Mel{\'e}ndez}, {Bean},
  {Ram{\'{\i}}rez}, {Leite}  \& {Asplund}}{{Bedell} et~al.}{2014}]{Bedell14}
{Bedell} M.,  {Mel{\'e}ndez} J.,  {Bean} J.~L.,  {Ram{\'{\i}}rez} I.,  {Leite}
  P.,   {Asplund} M.,  2014, \mn@doi [\apj] {10.1088/0004-637X/795/1/23}, \href
  {http://adsabs.harvard.edu/abs/2014ApJ...795...23B} {795, 23}

\bibitem[\protect\citeauthoryear{{Bedell} et~al.,}{{Bedell}
  et~al.}{2015}]{Bedell15}
{Bedell} M.,  et~al., 2015, \mn@doi [\aap] {10.1051/0004-6361/201525748}, \href
  {http://adsabs.harvard.edu/abs/2015A%26A...581A..34B} {581, A34}

\bibitem[\protect\citeauthoryear{{Bekki} \& {Tsujimoto}}{{Bekki} \&
  {Tsujimoto}}{2017}]{Bekki17}
{Bekki} K.,  {Tsujimoto} T.,  2017, \mn@doi [\apj] {10.3847/1538-4357/aa77ae},
  \href {http://adsabs.harvard.edu/abs/2017ApJ...844...34B} {844, 34}

\bibitem[\protect\citeauthoryear{{Bensby}, {Zenn}, {Oey}  \&
  {Feltzing}}{{Bensby} et~al.}{2007}]{Bensby07}
{Bensby} T.,  {Zenn} A.~R.,  {Oey} M.~S.,   {Feltzing} S.,  2007, \mn@doi
  [\apjl] {10.1086/519792}, \href
  {http://adsabs.harvard.edu/abs/2007ApJ...663L..13B} {663, L13}

\bibitem[\protect\citeauthoryear{{Bensby}, {Feltzing}  \& {Oey}}{{Bensby}
  et~al.}{2014}]{Bensby14}
{Bensby} T.,  {Feltzing} S.,   {Oey} M.~S.,  2014, \mn@doi [\aap]
  {10.1051/0004-6361/201322631}, \href
  {http://adsabs.harvard.edu/abs/2014A%26A...562A..71B} {562, A71}

\bibitem[\protect\citeauthoryear{{Biazzo} et~al.,}{{Biazzo}
  et~al.}{2015}]{Biazzo15}
{Biazzo} K.,  et~al., 2015, \mn@doi [\aap] {10.1051/0004-6361/201526375}, \href
  {http://adsabs.harvard.edu/abs/2015A%26A...583A.135B} {583, A135}

\bibitem[\protect\citeauthoryear{{Bisterzo}, {Travaglio}, {Gallino}, {Wiescher}
   \& {K{\"a}ppeler}}{{Bisterzo} et~al.}{2014}]{Bisterzo14}
{Bisterzo} S.,  {Travaglio} C.,  {Gallino} R.,  {Wiescher} M.,   {K{\"a}ppeler}
  F.,  2014, \mn@doi [\apj] {10.1088/0004-637X/787/1/10}, \href
  {http://adsabs.harvard.edu/abs/2014ApJ...787...10B} {787, 10}

\bibitem[\protect\citeauthoryear{{Busso}, {Gallino}  \& {Wasserburg}}{{Busso}
  et~al.}{1999}]{Busso99}
{Busso} M.,  {Gallino} R.,   {Wasserburg} G.~J.,  1999, \mn@doi [\araa]
  {10.1146/annurev.astro.37.1.239}, \href
  {http://adsabs.harvard.edu/abs/1999ARA%26A..37..239B} {37, 239}

\bibitem[\protect\citeauthoryear{{Busso}, {Marengo}, {Travaglio}, {Corcione}
  \& {Silvestro}}{{Busso} et~al.}{2001}]{Busso01}
{Busso} M.,  {Marengo} M.,  {Travaglio} C.,  {Corcione} L.,   {Silvestro} G.,
  2001, \memsai, \href {http://adsabs.harvard.edu/abs/2001MmSAI..72..309B} {72,
  309}

\bibitem[\protect\citeauthoryear{{Castelli} \& {Kurucz}}{{Castelli} \&
  {Kurucz}}{2004}]{Castelli04}
{Castelli} F.,  {Kurucz} R.~L.,  2004, ArXiv Astrophysics e-prints, \href
  {http://adsabs.harvard.edu/abs/2004astro.ph..5087C} {}

\bibitem[\protect\citeauthoryear{{Chiappini}, {Matteucci}  \&
  {Gratton}}{{Chiappini} et~al.}{1997}]{Chiappini97}
{Chiappini} C.,  {Matteucci} F.,   {Gratton} R.,  1997, \apj, \href
  {http://adsabs.harvard.edu/abs/1997ApJ...477..765C} {477, 765}

\bibitem[\protect\citeauthoryear{{Clayton}}{{Clayton}}{1988}]{Clayton88}
{Clayton} D.~D.,  1988, \mn@doi [\mnras] {10.1093/mnras/234.1.1}, \href
  {http://adsabs.harvard.edu/abs/1988MNRAS.234....1C} {234, 1}

\bibitem[\protect\citeauthoryear{{Connelly}, {Amelin}, {Krot}  \&
  {Bizzarro}}{{Connelly} et~al.}{2008}]{Connelly08}
{Connelly} J.~N.,  {Amelin} Y.,  {Krot} A.~N.,   {Bizzarro} M.,  2008, \mn@doi
  [\apjl] {10.1086/533586}, \href
  {http://adsabs.harvard.edu/abs/2008ApJ...675L.121C} {675, L121}

\bibitem[\protect\citeauthoryear{{C{\^o}t{\'e}}, {Belczynski}, {Fryer},
  {Ritter}, {Paul}, {Wehmeyer}  \& {O'Shea}}{{C{\^o}t{\'e}}
  et~al.}{2017}]{Cote17}
{C{\^o}t{\'e}} B.,  {Belczynski} K.,  {Fryer} C.~L.,  {Ritter} C.,  {Paul} A.,
  {Wehmeyer} B.,   {O'Shea} B.~W.,  2017, \mn@doi [\apj]
  {10.3847/1538-4357/aa5c8d}, \href
  {http://adsabs.harvard.edu/abs/2017ApJ...836..230C} {836, 230}

\bibitem[\protect\citeauthoryear{{Cowan} \& {Sneden}}{{Cowan} \&
  {Sneden}}{2004}]{Cowan04}
{Cowan} J.~J.,  {Sneden} C.,  2004, Origin and Evolution of the Elements, \href
  {http://adsabs.harvard.edu/abs/2004oee..symp...27C} {p.~27}

\bibitem[\protect\citeauthoryear{{Cox}}{{Cox}}{2000}]{Cox00}
{Cox} A.~N.,  2000, {Introduction}.
p.~1

\bibitem[\protect\citeauthoryear{{Cristallo}, {Straniero}, {Gallino},
  {Piersanti}, {Dom{\'{\i}}nguez}  \& {Lederer}}{{Cristallo}
  et~al.}{2009}]{Cristallo09}
{Cristallo} S.,  {Straniero} O.,  {Gallino} R.,  {Piersanti} L.,
  {Dom{\'{\i}}nguez} I.,   {Lederer} M.~T.,  2009, \mn@doi [\apj]
  {10.1088/0004-637X/696/1/797}, \href
  {http://adsabs.harvard.edu/abs/2009ApJ...696..797C} {696, 797}

\bibitem[\protect\citeauthoryear{{Cristallo}, {Straniero}, {Piersanti}  \&
  {Gobrecht}}{{Cristallo} et~al.}{2015}]{Cristallo15}
{Cristallo} S.,  {Straniero} O.,  {Piersanti} L.,   {Gobrecht} D.,  2015,
  \mn@doi [\apjs] {10.1088/0067-0049/219/2/40}, \href
  {http://adsabs.harvard.edu/abs/2015ApJS..219...40C} {219, 40}

\bibitem[\protect\citeauthoryear{{Cunha} \& {Lambert}}{{Cunha} \&
  {Lambert}}{1992}]{Cunha92}
{Cunha} K.,  {Lambert} D.~L.,  1992, \mn@doi [\apj] {10.1086/171950}, \href
  {http://adsabs.harvard.edu/abs/1992ApJ...399..586C} {399, 586}

\bibitem[\protect\citeauthoryear{{D'Orazi}, {Magrini}, {Randich}, {Galli},
  {Busso}  \& {Sestito}}{{D'Orazi} et~al.}{2009}]{DOrazi09c}
{D'Orazi} V.,  {Magrini} L.,  {Randich} S.,  {Galli} D.,  {Busso} M.,
  {Sestito} P.,  2009, \mn@doi [\apjl] {10.1088/0004-637X/693/1/L31}, \href
  {http://adsabs.harvard.edu/abs/2009ApJ...693L..31D} {693, L31}

\bibitem[\protect\citeauthoryear{{D'Orazi}, {Biazzo}, {Desidera}, {Covino},
  {Andrievsky}  \& {Gratton}}{{D'Orazi} et~al.}{2012}]{DOrazi12}
{D'Orazi} V.,  {Biazzo} K.,  {Desidera} S.,  {Covino} E.,  {Andrievsky} S.~M.,
   {Gratton} R.~G.,  2012, \mn@doi [\mnras] {10.1111/j.1365-2966.2012.21088.x},
  \href {http://adsabs.harvard.edu/abs/2012MNRAS.423.2789D} {423, 2789}

\bibitem[\protect\citeauthoryear{{D'Orazi}, {De Silva}  \& {Melo}}{{D'Orazi}
  et~al.}{2017}]{DOrazi17}
{D'Orazi} V.,  {De Silva} G.~M.,   {Melo} C.~F.~H.,  2017, \mn@doi [\aap]
  {10.1051/0004-6361/201629888}, \href
  {http://adsabs.harvard.edu/abs/2017A%26A...598A..86D} {598, A86}

\bibitem[\protect\citeauthoryear{{De Silva}, {D'Orazi}, {Melo}, {Torres},
  {Gieles}, {Quast}  \& {Sterzik}}{{De Silva} et~al.}{2013}]{DeSilva13}
{De Silva} G.~M.,  {D'Orazi} V.,  {Melo} C.,  {Torres} C.~A.~O.,  {Gieles} M.,
  {Quast} G.~R.,   {Sterzik} M.,  2013, \mn@doi [\mnras]
  {10.1093/mnras/stt153}, \href
  {http://adsabs.harvard.edu/abs/2013MNRAS.431.1005D} {431, 1005}

\bibitem[\protect\citeauthoryear{{Delgado Mena}, {Tsantaki}, {Adibekyan},
  {Sousa}, {Santos}, {Gonz{\'a}lez Hern{\'a}ndez}  \& {Israelian}}{{Delgado
  Mena} et~al.}{2017}]{DelgadoMena17}
{Delgado Mena} E.,  {Tsantaki} M.,  {Adibekyan} V.~Z.,  {Sousa} S.~G.,
  {Santos} N.~C.,  {Gonz{\'a}lez Hern{\'a}ndez} J.~I.,   {Israelian} G.,  2017,
  preprint, \href {http://adsabs.harvard.edu/abs/2017arXiv170504349D} {}
  (\mn@eprint {arXiv} {1705.04349})

\bibitem[\protect\citeauthoryear{{Dotter}, {Conroy}, {Cargile}  \&
  {Asplund}}{{Dotter} et~al.}{2017}]{Dotter17}
{Dotter} A.,  {Conroy} C.,  {Cargile} P.,   {Asplund} M.,  2017, \mn@doi [\apj]
  {10.3847/1538-4357/aa6d10}, \href
  {http://adsabs.harvard.edu/abs/2017ApJ...840...99D} {840, 99}

\bibitem[\protect\citeauthoryear{{Epstein}, {Johnson}, {Dong}, {Udalski},
  {Gould}  \& {Becker}}{{Epstein} et~al.}{2010}]{Epstein10}
{Epstein} C.~R.,  {Johnson} J.~A.,  {Dong} S.,  {Udalski} A.,  {Gould} A.,
  {Becker} G.,  2010, \mn@doi [\apj] {10.1088/0004-637X/709/1/447}, \href
  {http://adsabs.harvard.edu/abs/2010ApJ...709..447E} {709, 447}

\bibitem[\protect\citeauthoryear{{Feltzing}, {Howes}, {McMillan}  \&
  {Stonkut{\.e}}}{{Feltzing} et~al.}{2017}]{Feltzing17}
{Feltzing} S.,  {Howes} L.~M.,  {McMillan} P.~J.,   {Stonkut{\.e}} E.,  2017,
  \mn@doi [\mnras] {10.1093/mnrasl/slw209}, \href
  {http://adsabs.harvard.edu/abs/2017MNRAS.465L.109F} {465, L109}

\bibitem[\protect\citeauthoryear{{Fishlock}, {Karakas}, {Lugaro}  \&
  {Yong}}{{Fishlock} et~al.}{2014}]{Fishlock14}
{Fishlock} C.~K.,  {Karakas} A.~I.,  {Lugaro} M.,   {Yong} D.,  2014, \mn@doi
  [\apj] {10.1088/0004-637X/797/1/44}, \href
  {http://adsabs.harvard.edu/abs/2014ApJ...797...44F} {797, 44}

\bibitem[\protect\citeauthoryear{{Gaia Collaboration} et~al.,}{{Gaia
  Collaboration} et~al.}{2016}]{Gaia16}
{Gaia Collaboration} et~al., 2016, \mn@doi [\aap]
  {10.1051/0004-6361/201629512}, \href
  {http://adsabs.harvard.edu/abs/2016A%26A...595A...2G} {595, A2}

\bibitem[\protect\citeauthoryear{{Gallino}, {Arlandini}, {Busso}, {Lugaro},
  {Travaglio}, {Straniero}, {Chieffi}  \& {Limongi}}{{Gallino}
  et~al.}{1998}]{Gallino98}
{Gallino} R.,  {Arlandini} C.,  {Busso} M.,  {Lugaro} M.,  {Travaglio} C.,
  {Straniero} O.,  {Chieffi} A.,   {Limongi} M.,  1998, \mn@doi [\apj]
  {10.1086/305437}, \href {http://adsabs.harvard.edu/abs/1998ApJ...497..388G}
  {497, 388}

\bibitem[\protect\citeauthoryear{{Gilmore}, {Wyse}  \& {Kuijken}}{{Gilmore}
  et~al.}{1989}]{Gilmore89}
{Gilmore} G.,  {Wyse} R.~F.~G.,   {Kuijken} K.,  1989, \mn@doi [\araa]
  {10.1146/annurev.aa.27.090189.003011}, \href
  {http://adsabs.harvard.edu/abs/1989ARA%26A..27..555G} {27, 555}

\bibitem[\protect\citeauthoryear{{Goriely} \& {Siess}}{{Goriely} \&
  {Siess}}{2004}]{Goriely04}
{Goriely} S.,  {Siess} L.,  2004, \mn@doi [\aap] {10.1051/0004-6361:20040184},
  \href {http://adsabs.harvard.edu/abs/2004A%26A...421L..25G} {421, L25}

\bibitem[\protect\citeauthoryear{{Haywood}, {Di Matteo}, {Lehnert}, {Katz}  \&
  {G{\'o}mez}}{{Haywood} et~al.}{2013}]{Haywood13}
{Haywood} M.,  {Di Matteo} P.,  {Lehnert} M.~D.,  {Katz} D.,   {G{\'o}mez} A.,
  2013, \mn@doi [\aap] {10.1051/0004-6361/201321397}, \href
  {http://adsabs.harvard.edu/abs/2013A%26A...560A.109H} {560, A109}

\bibitem[\protect\citeauthoryear{{Haywood}, {Di Matteo}, {Snaith}  \&
  {Lehnert}}{{Haywood} et~al.}{2015}]{Haywood15}
{Haywood} M.,  {Di Matteo} P.,  {Snaith} O.,   {Lehnert} M.~D.,  2015, \mn@doi
  [\aap] {10.1051/0004-6361/201425459}, \href
  {http://adsabs.harvard.edu/abs/2015A%26A...579A...5H} {579, A5}

\bibitem[\protect\citeauthoryear{{Haywood}, {Lehnert}, {Di Matteo}, {Snaith},
  {Schultheis}, {Katz}  \& {G{\'o}mez}}{{Haywood} et~al.}{2016}]{Haywood16}
{Haywood} M.,  {Lehnert} M.~D.,  {Di Matteo} P.,  {Snaith} O.,  {Schultheis}
  M.,  {Katz} D.,   {G{\'o}mez} A.,  2016, \mn@doi [\aap]
  {10.1051/0004-6361/201527567}, \href
  {http://adsabs.harvard.edu/abs/2016A%26A...589A..66H} {589, A66}

\bibitem[\protect\citeauthoryear{{Holmberg}, {Nordstr{\"o}m}  \&
  {Andersen}}{{Holmberg} et~al.}{2009}]{Holmberg09}
{Holmberg} J.,  {Nordstr{\"o}m} B.,   {Andersen} J.,  2009, \mn@doi [\aap]
  {10.1051/0004-6361/200811191}, \href
  {http://adsabs.harvard.edu/abs/2009A%26A...501..941H} {501, 941}

\bibitem[\protect\citeauthoryear{{Karakas} \& {Lattanzio}}{{Karakas} \&
  {Lattanzio}}{2014}]{Karakas14}
{Karakas} A.~I.,  {Lattanzio} J.~C.,  2014, \mn@doi [\pasa]
  {10.1017/pasa.2014.21}, \href
  {http://adsabs.harvard.edu/abs/2014PASA...31...30K} {31, e030}

\bibitem[\protect\citeauthoryear{{Karakas} \& {Lugaro}}{{Karakas} \&
  {Lugaro}}{2016}]{Karakas16}
{Karakas} A.~I.,  {Lugaro} M.,  2016, \apj, submitted

\bibitem[\protect\citeauthoryear{{Kharchenko}, {Piskunov}, {Schilbach},
  {Roeser}, {Scholz}  \& {Zinnecker}}{{Kharchenko} et~al.}{2009}]{Kharchenko09}
{Kharchenko} N.~V.,  {Piskunov} A.~E.,  {Schilbach} E.,  {Roeser} S.,  {Scholz}
  R.-D.,   {Zinnecker} H.,  2009, VizieR Online Data Catalog, \href
  {http://adsabs.harvard.edu/abs/2009yCat..35040681K} {350, 40681}

\bibitem[\protect\citeauthoryear{{Kim}, {Demarque}, {Yi}  \& {Alexander}}{{Kim}
  et~al.}{2002}]{Kim02}
{Kim} Y.-C.,  {Demarque} P.,  {Yi} S.~K.,   {Alexander} D.~R.,  2002, \mn@doi
  [\apjs] {10.1086/343041}, \href
  {http://adsabs.harvard.edu/abs/2002ApJS..143..499K} {143, 499}

\bibitem[\protect\citeauthoryear{{Kordopatis} et~al.,}{{Kordopatis}
  et~al.}{2015}]{Kordopatis15}
{Kordopatis} G.,  et~al., 2015, \mn@doi [\aap] {10.1051/0004-6361/201526258},
  \href {http://adsabs.harvard.edu/abs/2015A%26A...582A.122K} {582, A122}

\bibitem[\protect\citeauthoryear{{Korobkin}, {Rosswog}, {Arcones}  \&
  {Winteler}}{{Korobkin} et~al.}{2012}]{Korobkin12}
{Korobkin} O.,  {Rosswog} S.,  {Arcones} A.,   {Winteler} C.,  2012, \mn@doi
  [\mnras] {10.1111/j.1365-2966.2012.21859.x}, \href
  {http://adsabs.harvard.edu/abs/2012MNRAS.426.1940K} {426, 1940}

\bibitem[\protect\citeauthoryear{{Kratz}, {Farouqi}, {Pfeiffer}, {Truran},
  {Sneden}  \& {Cowan}}{{Kratz} et~al.}{2007}]{Kratz07}
{Kratz} K.-L.,  {Farouqi} K.,  {Pfeiffer} B.,  {Truran} J.~W.,  {Sneden} C.,
  {Cowan} J.~J.,  2007, \mn@doi [\apj] {10.1086/517495}, \href
  {http://adsabs.harvard.edu/abs/2007ApJ...662...39K} {662, 39}

\bibitem[\protect\citeauthoryear{{Lachaume}, {Dominik}, {Lanz}  \&
  {Habing}}{{Lachaume} et~al.}{1999}]{Lachaume99}
{Lachaume} R.,  {Dominik} C.,  {Lanz} T.,   {Habing} H.~J.,  1999, \aap, \href
  {http://adsabs.harvard.edu/abs/1999A%26A...348..897L} {348, 897}

\bibitem[\protect\citeauthoryear{{Lallement}, {Vergely}, {Valette},
  {Puspitarini}, {Eyer}  \& {Casagrande}}{{Lallement}
  et~al.}{2014}]{Lallement14}
{Lallement} R.,  {Vergely} J.-L.,  {Valette} B.,  {Puspitarini} L.,  {Eyer} L.,
    {Casagrande} L.,  2014, \mn@doi [\aap] {10.1051/0004-6361/201322032}, \href
  {http://adsabs.harvard.edu/abs/2014A%26A...561A..91L} {561, A91}

\bibitem[\protect\citeauthoryear{{Lugaro}, {Herwig}, {Lattanzio}, {Gallino}  \&
  {Straniero}}{{Lugaro} et~al.}{2003}]{Lugaro03}
{Lugaro} M.,  {Herwig} F.,  {Lattanzio} J.~C.,  {Gallino} R.,   {Straniero} O.,
   2003, \mn@doi [\apj] {10.1086/367887}, \href
  {http://adsabs.harvard.edu/abs/2003ApJ...586.1305L} {586, 1305}

\bibitem[\protect\citeauthoryear{{Lugaro}, {Karakas}, {Stancliffe}  \&
  {Rijs}}{{Lugaro} et~al.}{2012}]{Lugaro12}
{Lugaro} M.,  {Karakas} A.~I.,  {Stancliffe} R.~J.,   {Rijs} C.,  2012, \mn@doi
  [\apj] {10.1088/0004-637X/747/1/2}, \href
  {http://adsabs.harvard.edu/abs/2012ApJ...747....2L} {747, 2}

\bibitem[\protect\citeauthoryear{{Magrini}, {Sestito}, {Randich}  \&
  {Galli}}{{Magrini} et~al.}{2009}]{Magrini09}
{Magrini} L.,  {Sestito} P.,  {Randich} S.,   {Galli} D.,  2009, \mn@doi [\aap]
  {10.1051/0004-6361:200810634}, \href
  {http://adsabs.harvard.edu/abs/2009A%26A...494...95M} {494, 95}

\bibitem[\protect\citeauthoryear{{Magrini} et~al.,}{{Magrini}
  et~al.}{2017}]{Magrini17}
{Magrini} L.,  et~al., 2017, \mn@doi [\aap] {10.1051/0004-6361/201630294},
  \href {http://adsabs.harvard.edu/abs/2017A%26A...603A...2M} {603, A2}

\bibitem[\protect\citeauthoryear{{Maiorca}, {Magrini}, {Busso}, {Randich},
  {Palmerini}  \& {Trippella}}{{Maiorca} et~al.}{2012}]{Maiorca12}
{Maiorca} E.,  {Magrini} L.,  {Busso} M.,  {Randich} S.,  {Palmerini} S.,
  {Trippella} O.,  2012, \mn@doi [\apj] {10.1088/0004-637X/747/1/53}, \href
  {http://adsabs.harvard.edu/abs/2012ApJ...747...53M} {747, 53}

\bibitem[\protect\citeauthoryear{{Maoz}, {Mannucci}  \& {Nelemans}}{{Maoz}
  et~al.}{2014}]{Maoz14}
{Maoz} D.,  {Mannucci} F.,   {Nelemans} G.,  2014, \mn@doi [\araa]
  {10.1146/annurev-astro-082812-141031}, \href
  {http://adsabs.harvard.edu/abs/2014ARA%26A..52..107M} {52, 107}

\bibitem[\protect\citeauthoryear{{Masseron} \& {Gilmore}}{{Masseron} \&
  {Gilmore}}{2015}]{Masseron15}
{Masseron} T.,  {Gilmore} G.,  2015, \mn@doi [\mnras] {10.1093/mnras/stv1731},
  \href {http://adsabs.harvard.edu/abs/2015MNRAS.453.1855M} {453, 1855}

\bibitem[\protect\citeauthoryear{{Matteucci}}{{Matteucci}}{2014}]{Matteucci14}
{Matteucci} F.,  2014, \mn@doi [The Origin of the Galaxy and Local Group,
  Saas-Fee Advanced Course, Volume 37.~ISBN 978-3-642-41719-1.~Springer-Verlag
  Berlin Heidelberg] {10.1007/978-3-642-41720-7_2}, \href
  {http://adsabs.harvard.edu/abs/2014SAAS...37..145M} {37, 145}

\bibitem[\protect\citeauthoryear{{Mayor} et~al.,}{{Mayor}
  et~al.}{2003}]{Mayor03}
{Mayor} M.,  et~al., 2003, The Messenger, \href
  {http://adsabs.harvard.edu/abs/2003Msngr.114...20M} {114, 20}

\bibitem[\protect\citeauthoryear{{Mel{\'e}ndez} et~al.,}{{Mel{\'e}ndez}
  et~al.}{2012}]{Melendez12}
{Mel{\'e}ndez} J.,  et~al., 2012, \mn@doi [\aap] {10.1051/0004-6361/201117222},
  \href {http://adsabs.harvard.edu/abs/2012A%26A...543A..29M} {543, A29}

\bibitem[\protect\citeauthoryear{{Mel{\'e}ndez} et~al.,}{{Mel{\'e}ndez}
  et~al.}{2014}]{Melendez14}
{Mel{\'e}ndez} J.,  et~al., 2014, \mn@doi [\apj] {10.1088/0004-637X/791/1/14},
  \href {http://adsabs.harvard.edu/abs/2014ApJ...791...14M} {791, 14}

\bibitem[\protect\citeauthoryear{{Mel{\'e}ndez} et~al.,}{{Mel{\'e}ndez}
  et~al.}{2017}]{Melendez17}
{Mel{\'e}ndez} J.,  et~al., 2017, \mn@doi [\aap] {10.1051/0004-6361/201527775},
  \href {http://adsabs.harvard.edu/abs/2017A%26A...597A..34M} {597, A34}

\bibitem[\protect\citeauthoryear{{Mennekens} \& {Vanbeveren}}{{Mennekens} \&
  {Vanbeveren}}{2016}]{Mennekens16}
{Mennekens} N.,  {Vanbeveren} D.,  2016, \mn@doi [\aap]
  {10.1051/0004-6361/201628193}, \href
  {http://adsabs.harvard.edu/abs/2016A%26A...589A..64M} {589, A64}

\bibitem[\protect\citeauthoryear{{Mishenina} et~al.,}{{Mishenina}
  et~al.}{2015}]{Mishenina15}
{Mishenina} T.,  et~al., 2015, \mn@doi [\mnras] {10.1093/mnras/stu2337}, \href
  {http://adsabs.harvard.edu/abs/2015MNRAS.446.3651M} {446, 3651}

\bibitem[\protect\citeauthoryear{{Nissen}}{{Nissen}}{2015}]{Nissen15}
{Nissen} P.~E.,  2015, \mn@doi [\aap] {10.1051/0004-6361/201526269}, \href
  {http://adsabs.harvard.edu/abs/2015A%26A...579A..52N} {579, A52}

\bibitem[\protect\citeauthoryear{{Nissen}}{{Nissen}}{2016}]{Nissen16}
{Nissen} P.~E.,  2016, preprint, \href
  {http://adsabs.harvard.edu/abs/2016arXiv160608399N} {} (\mn@eprint {arXiv}
  {1606.08399})

\bibitem[\protect\citeauthoryear{{Nomoto}, {Kobayashi}  \& {Tominaga}}{{Nomoto}
  et~al.}{2013}]{Nomoto13}
{Nomoto} K.,  {Kobayashi} C.,   {Tominaga} N.,  2013, \mn@doi [\araa]
  {10.1146/annurev-astro-082812-140956}, \href
  {http://adsabs.harvard.edu/abs/2013ARA%26A..51..457N} {51, 457}

\bibitem[\protect\citeauthoryear{{Nordstr{\"o}m} et~al.,}{{Nordstr{\"o}m}
  et~al.}{2004}]{Nordstrom04}
{Nordstr{\"o}m} B.,  et~al., 2004, \mn@doi [\aap] {10.1051/0004-6361:20035959},
  \href {http://adsabs.harvard.edu/abs/2004A%26A...418..989N} {418, 989}

\bibitem[\protect\citeauthoryear{{Pignatari} et~al.,}{{Pignatari}
  et~al.}{2016}]{Pignatari16}
{Pignatari} M.,  et~al., 2016, \mn@doi [\apjs] {10.3847/0067-0049/225/2/24},
  \href {http://adsabs.harvard.edu/abs/2016ApJS..225...24P} {225, 24}

\bibitem[\protect\citeauthoryear{{Ram{\'{\i}}rez}, {Mel{\'e}ndez}  \&
  {Asplund}}{{Ram{\'{\i}}rez} et~al.}{2014a}]{Ramirez14}
{Ram{\'{\i}}rez} I.,  {Mel{\'e}ndez} J.,   {Asplund} M.,  2014a, \mn@doi [\aap]
  {10.1051/0004-6361/201322558}, \href
  {http://adsabs.harvard.edu/abs/2014A%26A...561A...7R} {561, A7}

\bibitem[\protect\citeauthoryear{{Ram{\'{\i}}rez} et~al.,}{{Ram{\'{\i}}rez}
  et~al.}{2014b}]{Ramirez14b}
{Ram{\'{\i}}rez} I.,  et~al., 2014b, \mn@doi [\aap]
  {10.1051/0004-6361/201424244}, \href
  {http://adsabs.harvard.edu/abs/2014A%26A...572A..48R} {572, A48}

\bibitem[\protect\citeauthoryear{{Recio-Blanco} et~al.,}{{Recio-Blanco}
  et~al.}{2014}]{RecioBlanco14}
{Recio-Blanco} A.,  et~al., 2014, \mn@doi [\aap] {10.1051/0004-6361/201322944},
  \href {http://adsabs.harvard.edu/abs/2014A%26A...567A...5R} {567, A5}

\bibitem[\protect\citeauthoryear{{Reddy} \& {Lambert}}{{Reddy} \&
  {Lambert}}{2015}]{Reddy15}
{Reddy} A.~B.~S.,  {Lambert} D.~L.,  2015, \mn@doi [\mnras]
  {10.1093/mnras/stv1876}, \href
  {http://adsabs.harvard.edu/abs/2015MNRAS.454.1976R} {454, 1976}

\bibitem[\protect\citeauthoryear{{Reddy} \& {Lambert}}{{Reddy} \&
  {Lambert}}{2017}]{Reddy17}
{Reddy} A.~B.~S.,  {Lambert} D.~L.,  2017, \mn@doi [\apj]
  {10.3847/1538-4357/aa81d6}, \href
  {http://adsabs.harvard.edu/abs/2017ApJ...845..151R} {845, 151}

\bibitem[\protect\citeauthoryear{{Reddy}, {Lambert}  \& {Allende
  Prieto}}{{Reddy} et~al.}{2006}]{Reddy06}
{Reddy} B.~E.,  {Lambert} D.~L.,   {Allende Prieto} C.,  2006, \mn@doi [\mnras]
  {10.1111/j.1365-2966.2006.10148.x}, \href
  {http://adsabs.harvard.edu/abs/2006MNRAS.367.1329R} {367, 1329}

\bibitem[\protect\citeauthoryear{{Rojas-Arriagada} et~al.,}{{Rojas-Arriagada}
  et~al.}{2016}]{RojasArriagada16}
{Rojas-Arriagada} A.,  et~al., 2016, \mn@doi [\aap]
  {10.1051/0004-6361/201526969}, \href
  {http://adsabs.harvard.edu/abs/2016A%26A...586A..39R} {586, A39}

\bibitem[\protect\citeauthoryear{{Rojas-Arriagada} et~al.,}{{Rojas-Arriagada}
  et~al.}{2017}]{RojasArriagada17}
{Rojas-Arriagada} A.,  et~al., 2017, \mn@doi [\aap]
  {10.1051/0004-6361/201629160}, \href
  {http://adsabs.harvard.edu/abs/2017A%26A...601A.140R} {601, A140}

\bibitem[\protect\citeauthoryear{{Rosswog}, {Korobkin}, {Arcones}, {Thielemann}
   \& {Piran}}{{Rosswog} et~al.}{2014}]{Rosswog14}
{Rosswog} S.,  {Korobkin} O.,  {Arcones} A.,  {Thielemann} F.-K.,   {Piran} T.,
   2014, \mn@doi [\mnras] {10.1093/mnras/stt2502}, \href
  {http://adsabs.harvard.edu/abs/2014MNRAS.439..744R} {439, 744}

\bibitem[\protect\citeauthoryear{{Salaris}, {Chieffi}  \&
  {Straniero}}{{Salaris} et~al.}{1993}]{Salaris93}
{Salaris} M.,  {Chieffi} A.,   {Straniero} O.,  1993, \mn@doi [\apj]
  {10.1086/173105}, \href {http://adsabs.harvard.edu/abs/1993ApJ...414..580S}
  {414, 580}

\bibitem[\protect\citeauthoryear{{Schirbel} et~al.,}{{Schirbel}
  et~al.}{2015}]{Schirbel15}
{Schirbel} L.,  et~al., 2015, \mn@doi [\aap] {10.1051/0004-6361/201527303},
  \href {http://adsabs.harvard.edu/abs/2015A%26A...584A.116S} {584, A116}

\bibitem[\protect\citeauthoryear{{Shen}, {Cooke}, {Ramirez-Ruiz}, {Madau},
  {Mayer}  \& {Guedes}}{{Shen} et~al.}{2015}]{Shen15}
{Shen} S.,  {Cooke} R.~J.,  {Ramirez-Ruiz} E.,  {Madau} P.,  {Mayer} L.,
  {Guedes} J.,  2015, \mn@doi [\apj] {10.1088/0004-637X/807/2/115}, \href
  {http://adsabs.harvard.edu/abs/2015ApJ...807..115S} {807, 115}

\bibitem[\protect\citeauthoryear{{Slumstrup}, {Grundahl}, {Brogaard},
  {Thygesen}, {Nissen}, {Jessen-Hansen}, {Van Eylen}  \&
  {Pedersen}}{{Slumstrup} et~al.}{2017}]{Slumstrup17}
{Slumstrup} D.,  {Grundahl} F.,  {Brogaard} K.,  {Thygesen} A.~O.,  {Nissen}
  P.~E.,  {Jessen-Hansen} J.,  {Van Eylen} V.,   {Pedersen} M.~G.,  2017,
  \mn@doi [\aap] {10.1051/0004-6361/201731492}, \href
  {http://adsabs.harvard.edu/abs/2017A%26A...604L...8S} {604, L8}

\bibitem[\protect\citeauthoryear{{Snaith}, {Haywood}, {Di Matteo}, {Lehnert},
  {Combes}, {Katz}  \& {G{\'o}mez}}{{Snaith} et~al.}{2015}]{Snaith15}
{Snaith} O.,  {Haywood} M.,  {Di Matteo} P.,  {Lehnert} M.~D.,  {Combes} F.,
  {Katz} D.,   {G{\'o}mez} A.,  2015, \mn@doi [\aap]
  {10.1051/0004-6361/201424281}, \href
  {http://adsabs.harvard.edu/abs/2015A%26A...578A..87S} {578, A87}

\bibitem[\protect\citeauthoryear{{Sneden}}{{Sneden}}{1973}]{Sneden73}
{Sneden} C.,  1973, \mn@doi [\apj] {10.1086/152374}, \href
  {http://adsabs.harvard.edu/abs/1973ApJ...184..839S} {184, 839}

\bibitem[\protect\citeauthoryear{{Sneden}, {Cowan}  \& {Gallino}}{{Sneden}
  et~al.}{2008}]{Sneden08}
{Sneden} C.,  {Cowan} J.~J.,   {Gallino} R.,  2008, \mn@doi [\araa]
  {10.1146/annurev.astro.46.060407.145207}, \href
  {http://adsabs.harvard.edu/abs/2008ARA%26A..46..241S} {46, 241}

\bibitem[\protect\citeauthoryear{{Spina}, {Mel{\'e}ndez}  \&
  {Ram{\'{\i}}rez}}{{Spina} et~al.}{2016a}]{Spina16}
{Spina} L.,  {Mel{\'e}ndez} J.,   {Ram{\'{\i}}rez} I.,  2016a, \mn@doi [\aap]
  {10.1051/0004-6361/201527429}, \href
  {http://adsabs.harvard.edu/abs/2016A%26A...585A.152S} {585, A152}

\bibitem[\protect\citeauthoryear{{Spina}, {Mel{\'e}ndez}, {Karakas},
  {Ram{\'{\i}}rez}, {Monroe}, {Asplund}  \& {Yong}}{{Spina}
  et~al.}{2016b}]{Spina16b}
{Spina} L.,  {Mel{\'e}ndez} J.,  {Karakas} A.~I.,  {Ram{\'{\i}}rez} I.,
  {Monroe} T.~R.,  {Asplund} M.,   {Yong} D.,  2016b, \mn@doi [\aap]
  {10.1051/0004-6361/201628557}, \href
  {http://adsabs.harvard.edu/abs/2016A%26A...593A.125S} {593, A125}

\bibitem[\protect\citeauthoryear{{Surman}, {McLaughlin}, {Ruffert}, {Janka}  \&
  {Hix}}{{Surman} et~al.}{2008}]{Surman08}
{Surman} R.,  {McLaughlin} G.~C.,  {Ruffert} M.,  {Janka} H.-T.,   {Hix} W.~R.,
   2008, \mn@doi [\apjl] {10.1086/589507}, \href
  {http://adsabs.harvard.edu/abs/2008ApJ...679L.117S} {679, L117}

\bibitem[\protect\citeauthoryear{{Thielemann} et~al.,}{{Thielemann}
  et~al.}{2011}]{Thielemann11}
{Thielemann} F.-K.,  et~al., 2011, \mn@doi [Progress in Particle and Nuclear
  Physics] {10.1016/j.ppnp.2011.01.032}, \href
  {http://adsabs.harvard.edu/abs/2011PrPNP..66..346T} {66, 346}

\bibitem[\protect\citeauthoryear{{Tokovinin}}{{Tokovinin}}{2014}]{Tokovinin14}
{Tokovinin} A.,  2014, \mn@doi [\aj] {10.1088/0004-6256/147/4/86}, \href
  {http://adsabs.harvard.edu/abs/2014AJ....147...86T} {147, 86}

\bibitem[\protect\citeauthoryear{{Tucci Maia}, {Ram{\'{\i}}rez},
  {Mel{\'e}ndez}, {Bedell}, {Bean}  \& {Asplund}}{{Tucci Maia}
  et~al.}{2016}]{TucciMaia16}
{Tucci Maia} M.,  {Ram{\'{\i}}rez} I.,  {Mel{\'e}ndez} J.,  {Bedell} M.,
  {Bean} J.~L.,   {Asplund} M.,  2016, \mn@doi [\aap]
  {10.1051/0004-6361/201527848}, \href
  {http://adsabs.harvard.edu/abs/2016A%26A...590A..32T} {590, A32}

\bibitem[\protect\citeauthoryear{{Vandenberg} \& {Bell}}{{Vandenberg} \&
  {Bell}}{1985}]{Vandenberg85}
{Vandenberg} D.~A.,  {Bell} R.~A.,  1985, \mn@doi [\apjs] {10.1086/191052},
  \href {http://adsabs.harvard.edu/abs/1985ApJS...58..561V} {58, 561}

\bibitem[\protect\citeauthoryear{{Wanajo}}{{Wanajo}}{2013}]{Wanajo13}
{Wanajo} S.,  2013, \mn@doi [\apjl] {10.1088/2041-8205/770/2/L22}, \href
  {http://adsabs.harvard.edu/abs/2013ApJ...770L..22W} {770, L22}

\bibitem[\protect\citeauthoryear{{Woosley} \& {Weaver}}{{Woosley} \&
  {Weaver}}{1995}]{Woosley95}
{Woosley} S.~E.,  {Weaver} T.~A.,  1995, \mn@doi [\apjs] {10.1086/192237},
  \href {http://adsabs.harvard.edu/abs/1995ApJS..101..181W} {101, 181}

\bibitem[\protect\citeauthoryear{{Yana Galarza}, {Mel{\'e}ndez},
  {Ram{\'{\i}}rez}, {Yong}, {Karakas}, {Asplund}  \& {Liu}}{{Yana Galarza}
  et~al.}{2016}]{YanaGalarza16a}
{Yana Galarza} J.,  {Mel{\'e}ndez} J.,  {Ram{\'{\i}}rez} I.,  {Yong} D.,
  {Karakas} A.~I.,  {Asplund} M.,   {Liu} F.,  2016, \mn@doi [\aap]
  {10.1051/0004-6361/201527912}, \href
  {http://adsabs.harvard.edu/abs/2016A%26A...589A..17Y} {589, A17}

\bibitem[\protect\citeauthoryear{{Yi}, {Demarque}, {Kim}, {Lee}, {Ree},
  {Lejeune}  \& {Barnes}}{{Yi} et~al.}{2001}]{Yi01}
{Yi} S.,  {Demarque} P.,  {Kim} Y.-C.,  {Lee} Y.-W.,  {Ree} C.~H.,  {Lejeune}
  T.,   {Barnes} S.,  2001, \mn@doi [\apjs] {10.1086/321795}, \href
  {http://adsabs.harvard.edu/abs/2001ApJS..136..417Y} {136, 417}

\bibitem[\protect\citeauthoryear{{da Silva}, {Porto de Mello}, {Milone}, {da
  Silva}, {Ribeiro}  \& {Rocha-Pinto}}{{da Silva} et~al.}{2012}]{DaSilva12}
{da Silva} R.,  {Porto de Mello} G.~F.,  {Milone} A.~C.,  {da Silva} L.,
  {Ribeiro} L.~S.,   {Rocha-Pinto} H.~J.,  2012, \mn@doi [\aap]
  {10.1051/0004-6361/201118751}, \href
  {http://adsabs.harvard.edu/abs/2012A%26A...542A..84D} {542, A84}

\bibitem[\protect\citeauthoryear{{dos Santos} et~al.,}{{dos Santos}
  et~al.}{2017}]{dosSantos17}
{dos Santos} L.~A.,  et~al., 2017, preprint, \href
  {http://adsabs.harvard.edu/abs/2017arXiv170807465D} {} (\mn@eprint {arXiv}
  {1708.07465})

\makeatother
\end{thebibliography}

%\begin{thebibliography}{99}
%\end{thebibliography}

%%%%%%%%%%%%%%%%%%%%%%%%%%%%%%%%%%%%%%%%%%%%%%%%%%

%%%%%%%%%%%%%%%%% APPENDICES %%%%%%%%%%%%%%%%%%%%%


%%%%%%%%%%%%%%%%%%%%%%%%%%%%%%%%%%%%%%%%%%%%%%%%%%

% Don't change these lines
\bsp	% typesetting comment
\label{lastpage}
\end{document}